\documentclass[11pt,preprint]{aastex}

\def\simlt{\lower.5ex\hbox{$\; \buildrel < \over \sim \;$}}

\usepackage{graphicx}
\usepackage{natbib}
\usepackage{epstopdf}
\usepackage{epsfig}
\usepackage{subfig}
\usepackage{color}
\usepackage{datetime}
\usepackage{bbding}

\begin{document}

\title{VLT Spectropolarimetry of Broad Absorption Line QSOs}
\author{M. A. DiPompeo\altaffilmark{1}, M. S. Brotherton\altaffilmark{1}, C. De Breuck\altaffilmark{2}}
\altaffiltext{1}{University of Wyoming, Dept. of Physics and Astronomy 3905, 1000 E. University, Laramie, WY 82071}
\altaffiltext{2}{European Southern Observatory, Karl Schwarzschild Strasse 2, 85748 Garching bei M\"{u}nchen, Germany}

\begin{abstract}
We present spectropolarimetry of 19 confirmed and 4 possible bright, southern broad absorption line (BAL) quasars from the European Southern Observatory (ESO) Very Large Telescope (VLT).  A wide range of redshifts is covered in the sample (from 0.9 to 3.4), and both low- and hi-ionization quasars are represented, as well as radio-loud and radio-quiet BALQSOs.  We continue to confirm previously established spectropolarimetric properties of BALQSOs, including the generally rising continuum polarization with shorter wavelengths and comparatively large fraction with high broad-band polarization (6 of 19 with polarizations $>2\%$).  Emission lines are polarized less than or similar to the continuum, except in a few unusual cases, and absorption troughs tend to have higher polarizations.  A search for correlations between polarization properties has been done, identifying 2 significant or marginally significant correlations.  These are an increase in continuum polarization with decreasing optical luminosity (increasing absolute B magnitude) and decreasing \ion{C}{4} emission-line polarization with increased continuum polarization.
\end{abstract}


\section{INTRODUCTION}

Around 20\% of optically selected quasars exhibit blueshifted, broad absorption lines (BALs) (e.g. Knigge et al. 2008), which indicate massive outflows from the quasar central engine that may have important consequences (Vernaleo \& Reynolds 2006, Hopkins et al. 2006, Scannapieco \& Oh 2004).  Outflows can enrich the quasar host galaxy ISM,  contribute to a variety of feedback effects such as regulating host galaxy star formation rates (Hopkins \& Elvis 2010), and limit quasar lifetimes by removing fuel from the nuclear regions (Silk \& Rees 1998, King 2003).  Despite their important implications, the nature of BALQSOs is still not fully understood.

BALQSOs are a highly polarized subclass of quasars, with 8 of 40 showing optical broadband polarization of 2\% or greater compared to 2 of 115 non-BALQSOs (Hines \& Schmidt 1997).  There have been a few spectropolarimetry surveys done in the last several years, for example Ogle et al. (1999) and Schmidt, Hines, and Smith (1997), and DiPompeo et al. (2010).  Spectropolarimetry of small samples and individual objects has been published as well, covering a wide range of BALQSO types (Goodrich \& Miller 1995, Cohen et al. 1995, Glenn et al. 1994).  In general, these results show common trends among polarized BALQSO properties.  Continuum polarization can exceed 10\% and the polarization generally increases at shorter wavelengths.  Polarization levels in absorption lines tend to be equal to or greater than that of the continuum and emission lines are generally similarly or less polarized than the continuum, though these are general trends and certainly not true of all BALQSOs.

A popular explanation for BALQSOs has involved a simple orientation scheme.  In this model all quasars have outflows, but only a fraction (about 20\%) are seen along a line of sight that intersects the outflow; these sources show BALs.  Different BALQSO subtypes, here described as HiBALs (those showing absorption from high-ionization species such as \ion{C}{4} $\lambda$1549\AA), LoBALs (those with additional absorption from low-ionization species such as \ion{Mg}{2} $\lambda$2799\AA), and FeLoBALS (LoBALs showing absorption from \ion{Fe}{2} $\lambda$2380\AA, $\lambda$2600\AA, $\lambda$2750\AA) are fit into this orientation paradigm as well.  LoBALs and FeLoBALs would be those BALQSOs seen at the largest inclinations.  This description of BALQSOs has relied heavily on analogies with the unification of Seyfert galaxies via orientation (Antonucci 1993).
The spectropolarimetric results have also often been taken to support the simple orientation model  (e.g. Goodrich \& Miller 1995, Hines \& Wills 1995, Cohen et al. 1995).  In recent years there has been growing evidence that the orientation explanation is at least incomplete, especially in its simplest forms.

BALQSOs with polar outflows (Zhou et al. 2006, Ghosh \& Punsly 2007) and the compactness of BALQSOs in radio maps (Montenegro-Montes et al. 2008, Becker et al. 2000) suggest that BALQSOs are not simply normal quasars seen edge-on.  A model based on evolution is another possibility, for example Gregg et al. (2002, 2006) suggest that the BAL phase is an early phase in the lifetime of all quasars, which would last for approximately 20\% of their lifetime.

Determining which of these models is correct, or if some combination of both is needed, is an important problem concerning our understanding of quasars.  To this end we have begun work on a project with a large sample of BALQSOs to investigate with spectropolarimetry, radio data, and optical data.  This paper is the second of several from the project (the first being DiPompeo et al. 2010), and our aim here is to present the spectropolarimetry from the European Southern Observatory Very Large Telescope (ESO VLT) sample, and report the most basic results.

We adopt the cosmology of Spergel et al. (2007) for all calculated properties, with $H_{0}=71$ km/s/Mpc, $\Omega_{M}=0.27$ and $\Omega_{\Lambda}=0.73$.

\section{TARGETS}
This set of objects was not selected to build a uniform sample across any particular range of properties.  Instead, many of these sources were chosen due to their extreme or unusual nature.  Studying the most unusual/extreme objects in a class is often exceptionally useful for shedding light on the physics that drives the extreme properties, and in turn those that are less extreme.  Many sources were simply chosen because they were bright enough to allow quick on-site reductions to identify polarized sources that would be worth giving a deeper look.  It should be kept in mind that this is a heterogenous sample, especially when looking at correlations.

The sample is made up of 19 BALQSOs and 4 sources that are uncertainly classified as BALQSOs, mostly found in the catalog of Sowinski et al. (1997).  Seven are from the list of unusual Sloan Digital Sky Survey (SDSS) BALQSOs in Hall et al. (2002); three of those are also from the southern installment of the First Bright Quasar Survey (FBQS, Becker et al. 2001).  Four sources are found in the Large Bright Quasar Survey (LBQS, Hewett et al. 1995).  Two objects are from a Very Large Array (VLA) survey of radio-selected (FIRST 1.4 GHz fluxes greater than 10 mJy) SDSS BALQSOs that we are currently conducting.  The remainder of the sources in the sample are simply bright ($16 \lesssim V \lesssim 20$) southern BALQSOs, with redshifts ranging from about 1.4 to 3.4.  

Of the seven sources found in Hall et al. (2002), four of them are uncertain in their classification as BALQSOs: FBQS J0105-0033, FBQS J2204+0031, FBQS J0127+0114, and SDSS J0338+0056.  The first of these two, FBQS J0105-0033 and FBQS J2204+0031, show no obvious absorption (the redshifts are too low for \ion{C}{4} to be seen in the spectrum), but may show \ion{Fe}{2} emission which correlates with the presence of BALs (Boroson \& Meyers 1992).  Both sources are highly reddened as well.  Becker et al. (2001) has classified these objects as BL Lacartae (BL Lac) sources, though they do note that the classification is uncertain.  FBQS J0127+0114 is similar to the previous two objects (and speculated to be a BAL for the same reasons), but it does also show small absorption from \ion{He}{1} $\lambda$3889\AA.  SDSS J0338+0056 is also highly reddened, which makes determining the continuum difficult to do.  Depending on where the continuum is defined, there may or may not be strong absorption from \ion{Mg}{2}, \ion{Al}{3}, \ion{Fe}{3}, and \ion{Fe}{2}.  With our higher S/N spectrum, we see no obvious absorption from \ion{C}{4}, though it is on the edge of the spectrum.  See Hall et al. (2002) for a more detailed discussion of these possible BALQSOs

Table~\ref{mainpropstbl} lists the targets with some basic properties.  Absolute $B$ magnitudes use the SDSS data release 7 (DR7) $g$-band magnitudes as a direct estimate of the apparent $B$ magnitude.  Not all of the targets are found in SDSS, however, and these are footnoted in the table.  A few sources had no published optical magnitudes, so $B$ magnitudes were measured from our total light spectra.  The ``class" column lists the BAL type (HiBAL, LoBAL, FeLoBAL).  The objects discussed above with uncertain classifications are followed by a ``?" in the table.

Six of the sources in this sample contain radio data in the literature; these are indicated in the final column of table~\ref{mainpropstbl}.  The radio properties of these sources are summarized in table~\ref{radiopropstbl}.  Columns 4 and 5 are the K-corrected rest-frame 5 GHz radio luminosity (log $L_r$) and radio-loudness parameter (log $R^{*}$), respectively, as defined in Stocke et al. (1992).  Five of the sources are considered formally radio-loud (log $R^{*}>1.0$), and the sixth is extremely close with log $R^{*}=0.9$.  The final column is the radio spectral index $\alpha_r$ ($f \propto \nu^{\alpha_r}$); unfortunately for all but one of the objects radio data is only available at one frequency.  It has thus been assumed for the calculations of $L_r$ and $R^{*}$ that $\alpha_r = -0.3$ (except for SDSS J0043-0015) and $\alpha_{opt} = -1$.  Only one of these radio sources shows extended structure, SDSS J0043-0015 (also known as PKS 0040-005).  This object has a double lobed FR II structure, with the lobes at a position angle of 51$\arcdeg$.  Brotherton et al. (2006) discussed the radio and polarimetric properties of this object, however it has since been discovered that the polarization position angle presented in that paper is incorrect due to a reduction error causing the position angle to be rotated by 90$\arcdeg$.  While the other points in the paper remain valid, the argument regarding the polarization and radio position angles appearing almost parallel is incorrect, and the angles are in fact closer to perpendicular (see \S 5 for more on this relationship).  An erratum is being submitted in parallel with this paper.

\section{SPECTROPOLARIMETRY}
\subsection{Observations}

Observations of 21 objects were performed on September 20 and 21, 2003, using the PMOS mode of the FORS1 spectrograph (Appenzeller et al. 1998) on the Antu unit of the VLT.  Conditions were photometric both nights.  Observations were typically an hour or less and broken into four exposures, one for each waveplate position (0$\arcdeg$, 45$\arcdeg$, 22.5$\arcdeg$, 67.5$\arcdeg$).  We used a 300 line mm$^{-1}$ grism blazed at 5000 \AA\ and a 1\arcsec\ slit, which gave an effective resolution of 10 \AA\ (FWHM of sky lines).  The dispersion was 2.6 \AA\ pixel$^{-1}$.

The two sources from our VLA survey were observed May 5 and 8, 2010, in VLT service mode with the PMOS mode of the FORS2 spectrograph (Appenzeller et al. 1998) on the Antu unit of the VLT.  Observations were approximately 1.4 hours total (all waveplate positions).  A 600 line mm$^{-1}$ grism blazed at 5000 \AA\ was used with a 1.4\arcsec\ slit, which gave an effective resolution of 10\AA\ (FWHM of sky lines) and a dispersion of 2 \AA\ pixel$^{-1}$.

\begin{figure}
 \centering
  \figurenum{1}
   \includegraphics[width=5in]{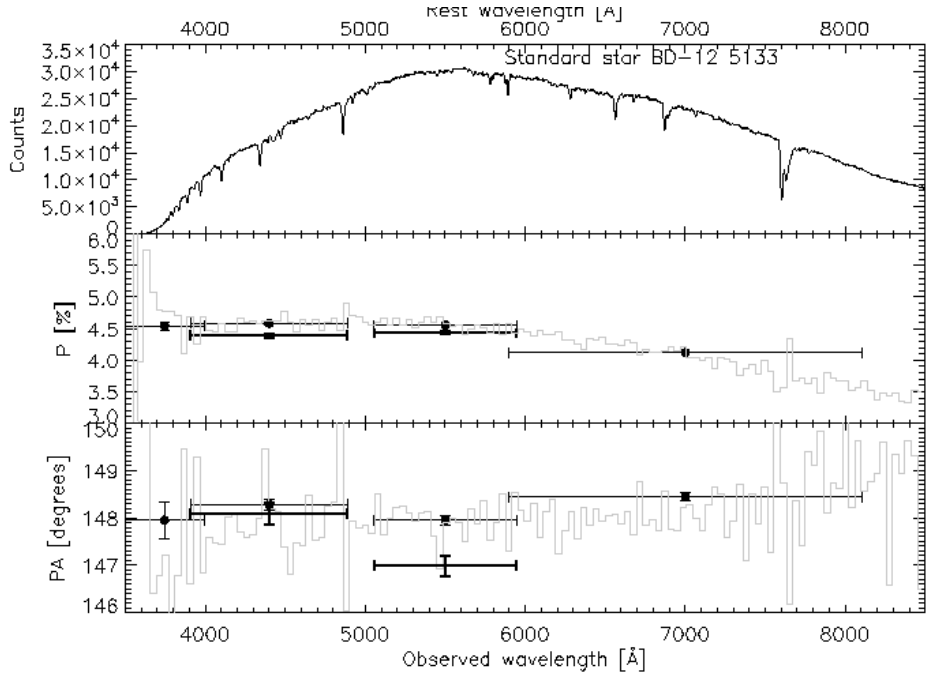}
  \caption{Spectropolarimetry results for the polarization standard star BD -12 5133.  The top panel is the total spectrum (not flux calibrated), the middle panel is the polarization level, and the bottom panel is the polarization position angle.  The polarization and position angle bins are standard UBVR wavelengths.  Bins marked in bold are B and V-band published values from the literature (Fossati et al. 2007).  Our values for the polarization $p$ are within about 0.1\%, and the position angle matches to within about 1$^{\circ}$.\label{stdstarfig}}
\end{figure}

The wavelength dependence of the polarization position angle due to the instrument optics was calibrated using ESO supplied data.  During the September 2003 observations standard stars\footnote{See \url{http://www.eso.org/sci/facilities/paranal/instruments/fors/inst/pola.html} for standard star information} BD-12-5133 and NGC 2024 NIR1 were observed both nights to check the polarization angle offset between the half-wave plate coordinates and the sky coordinates, as well as check the polarization percentages.  An example of the standard star results is shown in Figure~\ref{stdstarfig}.  Our polarization levels are accurate to within about 0.1\%, and the polarization position angle is accurate to about 1$\arcdeg$.  Additionally, the unpolarized standard star GD 50 was observed to ensure there was no significant instrumental polarization; levels were always less than 0.1\% for this star.  For the observations in May 2010, which were done in service mode, archival ESO data of the standard stars Vela1 95 and HDE 316232 taken nearby in time and with an identical instrument configuration were used; our polarization levels were again accurate to about 0.1\% and the position angle was accurate to 1$\arcdeg$ or less.  The unpolarized standard WD-1620-391 was observed as well and showed polarizations of 0.04\% or less.

\subsection{Measurements}

We reduced our data to one-dimensional spectra using standard techniques within
the IRAF NOAO package. The rms uncertainties in the dispersion solution were
0.2\AA, and we used sky lines to ensure that our zero point was accurate to
0.1\AA.  Wavelengths are air wavelengths.  We followed standard procedures
(Miller et al. 1988; Cohen et al. 1997) for calculating
Stokes parameters.  Polarizations reported are debiased using the rotated Stokes Q and U parameters, and uncertainties are 1$\sigma$ confidence intervals (Simmons \& Stewart 1985).  Binning is always done in flux before calculating Stokes parameters.

The spectropolarimetry results for the whole sample are plotted in Figures~\ref{polplotsfig} (a)-(w), showing the total flux spectrum, polarized flux ($p$ times $f$), linear polarization, and polarization position angle as a function of both rest-frame and observed-frame wavelengths.  Table~\ref{contpoltbl} lists the continuum and white-light (white-light measurements are averaged over the whole spectrum, including both lines and continuum) polarization properties of the sample.  Continuum polarization measurements were made between the \ion{C}{4} and \ion{C}{3}]/\ion{Al}{3} emission lines when possible because of the lack of other prominent features in this wavelength region.  If this region was not available, continuum measurements were made by subjectively choosing the cleanest part of the spectrum.  The sixth column indicates the wavelengths (observed frame) over which continuum measurements were made; these regions are also plotted in bold in the polarization panel of Figures~\ref{polplotsfig} (a)-(w).  The final column is the maximum interstellar polarization (max ISP) possible along the source's line of sight, calculated by taking $9\times E(B-V)$ (Serkowsky et al. 1975).  The $E(B-V)$ values are from Schlegel et al. (1998) via NED\footnote{This research has made use of the NASA/IPAC Extragalactic Database (NED) which is operated by the Jet Propulsion Laboratory, California Institute of Technology, under contract with the National Aeronautics and Space Administration.}.

In addition to continuum and white-light polarization measurements, we have also measured polarization properties in several emission and absorption features.  Table~\ref{linepoltbl} lists these measurements.  Average observed polarization in three main emission features (\ion{C}{4}, \ion{C}{3}], and \ion{Mg}{2}) are given in columns 2, 3, and 4 and are labeled as $p_e$.  BAL troughs of \ion{C}{4} and \ion{Mg}{2} were also analyzed; these values are given in columns 5 and 6, and are labeled as $p_a$.  Line measurements were made by binning across the features, only including the peak of emission or bottom of absorption troughs when possible.  However, in many instances wider bins that included the wings of the lines (but not any adjacent continuum) were needed in order to contain enough flux to make accurate measurements.  The last 5 columns of the table list line polarizations normalized by the polarization of the continuum immediately redward of the features, in order to account for the fact that polarization is generally a function of wavelength.  Entries of ``..." indicate that the feature was not present in the spectrum.  In some cases the signal-to-noise ratio or the polarization in the lines was only high enough to obtain upper limits.  These values are preceded by ``$<$" in the table.  Upper limits were not included in any subsequent analysis.

\subsection{Notes on Individual Objects}

\noindent \textbf{\textit{SDSS J2215-0045}}

\noindent This source has a continuum polarization measurement of 0.30\%, and a maximum ISP of 0.95\%, indicating that it is not likely intrinsically polarized.  Nowhere in the spectrum does the polarization rise above 1\%, and the position angle does not convincingly show any rotation through any spectral features.  This object is at best very weakly polarized, but is probably intrinsically unpolarized.

\noindent \textbf{\textit{FBQS J0105-0033, FBQS J2204+0031, \& SDSS J0338+0056}}

\noindent These three sources are uncertainly classified as BALs (see \S 2), and none appear to be convincingly intrinsically polarized.  FBQS J0105-0033 has a continuum polarization of 0.31\% and a maximum ISP of 0.32\%, and FBQS J2204+0031 has $p=0.22\%$ and a maximum ISP of 0.50\%.  We measure $p=0.71\%$ in SDSS J0338+0056, while the maximum ISP along its line of sight is 0.96\%.  The two FBQS sources do not show any significant rise in polarization throughout the spectrum, though the position angle is quite variable.  The SDSS source does have a few points where the polarization rises, in fact to over 2\% in the \ion{Mg}{2} emission line, though it is not extremely convincing.  The fact that these sources seem intrinsically unpolarized (or only very weakly polarized) may also be further evidence that they are not actual BALQSOs.  Interestingly, if they are in fact BALQSOs, they are all highly reddened yet do not show significant polarization; generally the most reddened sources in this class tend to be the most polarized.

\section{ANALYSIS \& RESULTS}
In the following analysis, in order to ensure that we are truly only seeing the properties of BALQSOs, we do not include the four sources discussed in \S2 for which the classification as BALQSOs is uncertain.  We also remind the reader that this is not a homogeneous sample of BALQSOs.

\begin{figure}
 \centering
  \figurenum{3}
   \includegraphics{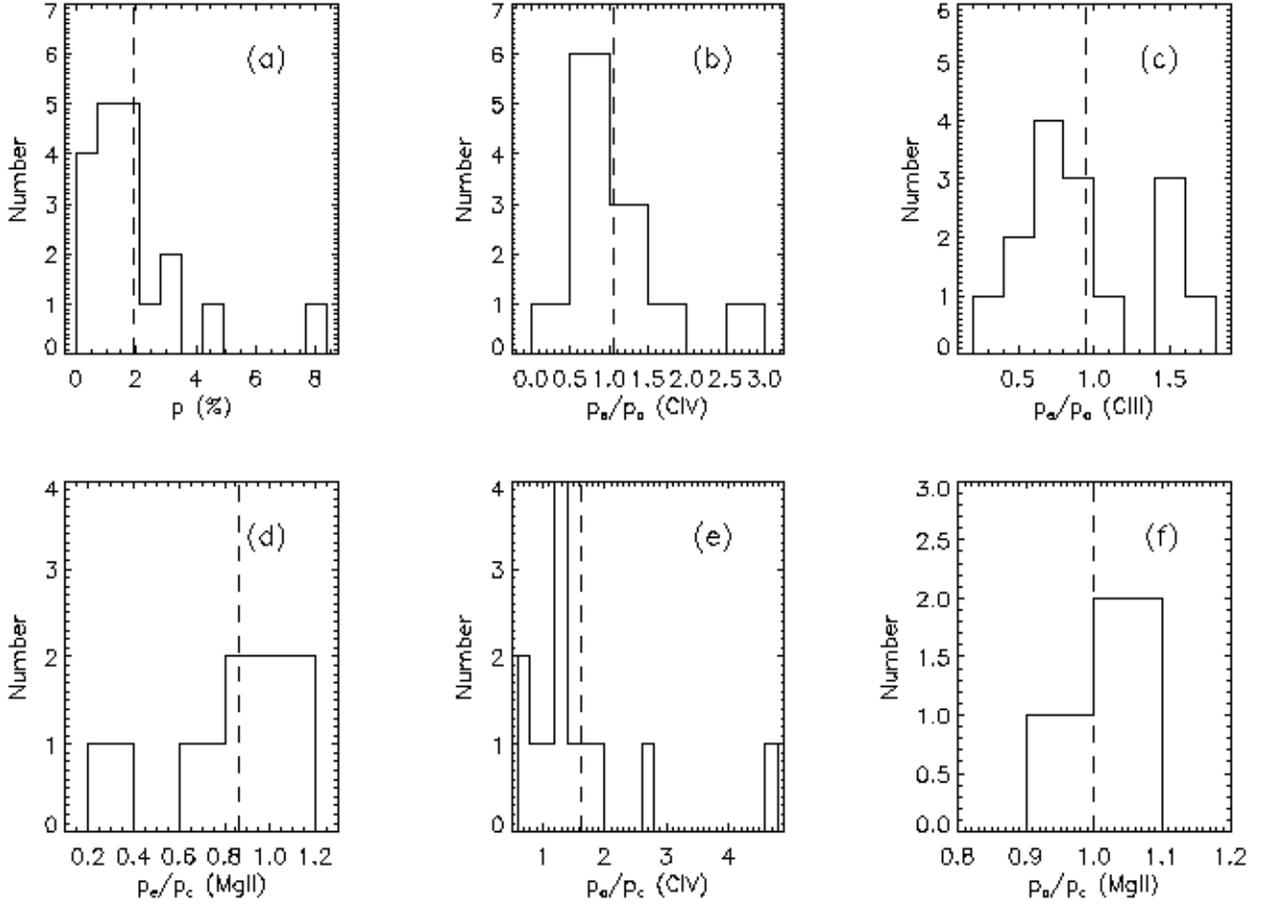}
  \caption{Histograms of the polarization properties of the sample.  Box (a) is the continuum polarization and boxes (b)-(d) are the emission line polarzations for \ion{C}{4}, \ion{C}{3}], and \ion{Mg}{2}, respectively, normalized by the adjacent continuum polarization.  Boxes (e) and (f) are the polarizations in the \ion{C}{4} and \ion{Mg}{2} absorption troughs (normalized by the adjacent continuum).  Vertical dashed lines indicate the mean of each property.\label{polplot}}
\end{figure}

Figure~\ref{polplot} shows the distributions in the polarization properties of the sample, with the mean of each property marked by a vertical dashed line.  The polarization in the emission lines generally seems consistent with or lower than the continuum.  The average values of $p_e/p_c$ in the \ion{C}{4}, \ion{C}{3}], and \ion{Mg}{2} are 1.06, 0.95, and 0.86, respectively.  However, we do note that there is one object, 2046-3433, that has unusually high polarization in the \ion{C}{4} emission line ($p_e/p_c =2.56$), and this is skewing the average for that line to higher values.  Not including this source gives an average value of 0.92.  Thus we can say that the emission lines are generally less polarized than the continuum.  We see for certain that the absorption lines do tend to be more polarized in the \ion{C}{4} line, with a mean $p_a/p_c$ of 1.60, and less certainly in the \ion{Mg}{2} absorption with a mean $p_a/p_c$ of 1.00.  We do note that given the width of these distributions and the number of sources, that many of these averages are not inconsistent with 1.

\begin{figure}
 \centering
   \figurenum{4}
     \includegraphics[width=6in]{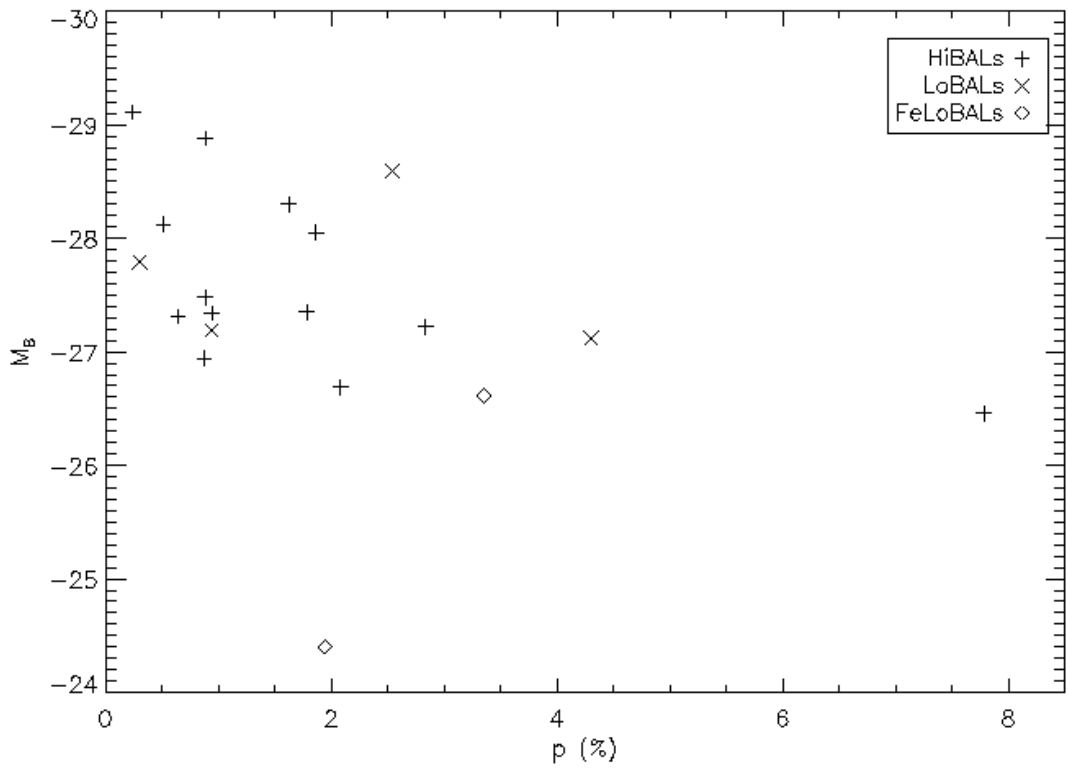}
    \caption{The correlation between absolute $B$ magnitude and continuum polarization.  As the luminosity of a BALQSO increases, it tends to be less polarized.\label{pmagcorr}}
\end{figure}

\begin{figure}
 \centering
   \figurenum{5}
     \includegraphics[width=6in]{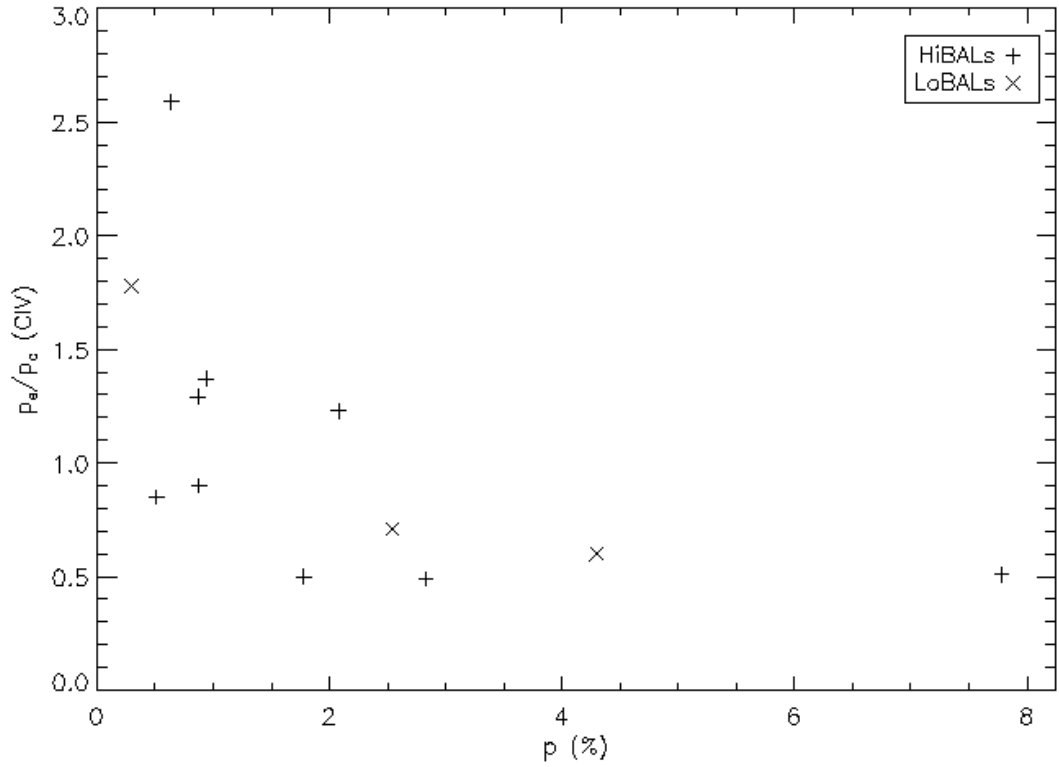}
    \caption{The correlation between continuum polarization and \ion{C}{4} emission-line polarization.  Recall that $p_c$ on the y-axis is the continuum polarization immediately to the red side of the emission line, and is not necessarilly the same as the continuum polarization on the x-axis.\label{ppecivcorr}}
\end{figure}

We have done a simple correlation test on the polarization properties using the Spearman $r_s$ rank correlation coefficient between property pairs that have at least 9 data points.  The results are shown in table~\ref{corrtbl}.  Given the small number of objects in the sample, we use a loose criterion of $P_{r_s} \le 0.05$ to indicate at least marginally significant correlations.  Values of $P_{r_s}$ satisfying this cutoff are shown in bold type in the table; while all property pairs are shown in the table for completeness, only those involving at least 9 sources were analyzed and will be considered further.  Any others require more data to verify.  Two correlations did satisfy our criteria: an anti-correlation between the continuum polarization and BALQSO luminosity (or a positive correlation with absolute $B$ magnitude, Figure~\ref{pmagcorr}) and an anti-correlation between continuum polarization and \ion{C}{4} emission polarization (Figure~\ref{ppecivcorr}).

\section{DISCUSSION}
The spectropolarimetric properties of this sample are consistent with previous findings.  If we define high broadband polarization as larger than 2\%, we find that 6 out of 19 ($32\% \pm 11\%$) BALQSOs in this sample are highly polarized (errors are 1$\sigma$ values from a binomial distribution).  This is higher than (but consistent with) the percentage in the radio-selected sample of DiPompeo et al. (2010) of $20\% \pm 7.3\%$, and the 8 of 40 ($20\% \pm 6\%$) reported by Hines \& Schmidt (1997). We also note that Hines \& Schmidt (1997) do not include objects for which $\sigma_p > 0.62\%$, and all of our sources satisfy that inequality.  Our data support the conclusion that between a quarter and one third of BALQSOs are highly polarized.  The general increase in polarization in absorption troughs and decrease in polarization in emission lines has also been consistently noted in BALQSO samples.  

There is one object however, QSO 2046-3433 that has high polarization in emission lines compared to the continuum.  This is most likely interpreted as a geometry in which the scattering region is blocked (possibly by lumps on the inner edge of a dusty torus) from the continuum source, while the broad-line region has a more clean line of sight to the scattering region.  While this does not necessarily lend support to any one BALQSO model, it seems unlikely that this would be seen in face-on BALQSOs.  This however only appears to occur in a very small number of sources.

\begin{figure}
 \centering
  \figurenum{6}
   \includegraphics[width=6in]{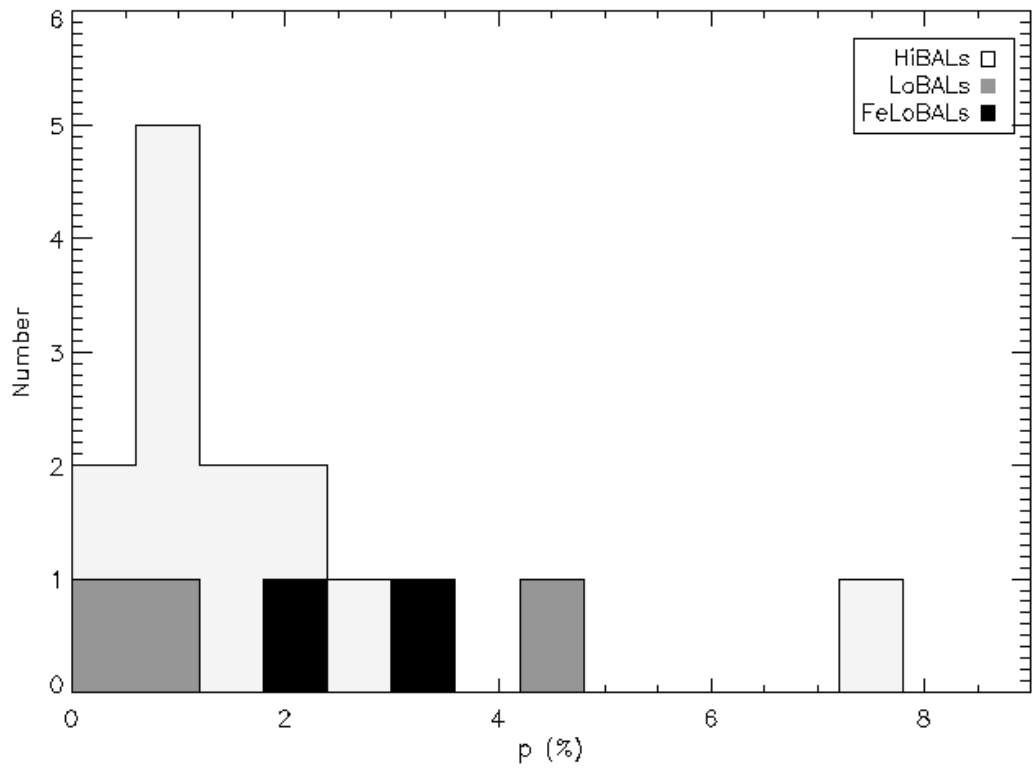}
  \caption{Comparison of the polarization distribution by BALQSO subtype.  No types completely cover another, so we have not plotted them separately.\label{polbytypefig}}  
\end{figure}

We have also compared the continuum polarizations for the different BALQSO types, shown in Figure~\ref{polbytypefig}.  It is not as clear as it has been in other samples that the LoBALs and FeLoBALs tend to be the most highly polarized BALQSOs.  This could be an indication that orientation is not the only factor in determining whether a quasar shows BALs or not, but we again note that the numbers are small and we will be able to analyze this more clearly when more objects are considered.

Of the correlations found, one has been previously identified- that between continuum polarization and absolute $B$ magnitude was also seen in DiPompeo et al. (2010).  There are two extreme points in this analysis, a HiBAL with very high polarization (almost 8\%) and an FeLoBAL with low luminosity ($M_B = -24.4$), and we have checked the correlation not including these points to examine its robustness to outliers.  Without the high polarization HiBAL, $P_{r_s}= 0.04$, and without the low luminosity FeLoBAL $P_{r_s}=0.02$, so the correlation remains significant.  If we exclude both extreme points, however, the value of $P_{r_s}$ becomes 0.07 and is no longer considered significant by our criteria.  Filling in the parameter space between the main group and the extreme points is necessary to determine if the correlation is truly significant.  We do not see the correlation between the \ion{C}{4} and \ion{C}{3}] emission-line polarizations, which was seen in both Lamy \& Hutsemekers (2004) and DiPompeo et al. (2010).  We defer analysis of these correlations and their implications to a later paper when we will be combining these data with that from DiPompeo et al. (2010), as well as new VLT observations of radio-selected BALQSOs.

Radio information in conjunction with polarization measurements can be useful in testing theories of BALQSOs (e.g. DiPompeo et al. 2010).  This works especially well for extended radio sources, which allows a comparison between the position angle of the radio jets (and thus the orientation of the BALQSO) and the polarization position angle.  In the one extended source in this sample, SDSS J0043-0015, the radio position angle is 51$\arcdeg$, while the polarization position angle is about 128$\arcdeg$ (both in white-light and in the continuum) The radio and polarization position angles are roughly perpendicular, which we would expect if the source is seen from an edge-on perspective.  However, the fact that only 1 of the 8 radio sources show extended structure could be support for evolutionary BALQSO models.

With only 6 radio sources in this data set, 3 of which are not certainly BALQSOs, statistical tests looking for correlations between radio and polarization properties would not be useful.  In the future we will however be combining these with the sample of DiPompeo et al. (2010) as well as new observations of radio-selected BALQSOs to provide a more significant analysis of the relationships between BALQSO radio properties and polarization.

\section{SUMMARY}
We present spectropolarimetry with the Very Large Telescope of 19 bright, southern BALQSOs as well as of 4 sources that may or may not be BALQSOs.  They cover a range of redshifts, from 0.9 to 3.4, and the various BAL subtypes are represented.  We confirm previous findings in polarimetric properties.  A large fraction of BALQSOs show high continuum polarization (32\% in this sample), continuum polarization tends to rise with shorter wavelengths, absorption troughs are generally more polarized than the continuum, and emission lines are generally less polarized.  Two correlations between properties are found; a negative correlation between continuum polarization and absolute $B$ magnitude, and an increase in continuum polarization with decreasing \ion{C}{4} emission-line polarization.  Subsequent papers will add more sources to these data in order to more accurately analyze these findings and place them in context with a consistent model for BALQSOs.

\acknowledgments

This project is based on observations collected at the European Southern Observatory, Paranal, projects 71.B-0121(A) and 85.B-0615(A).  We acknowledge support from NASA through grant \#NNG05GE84G and the Wyoming NASA Space Grant Consortium, NASA Grant \#NNG05G165H.  M. DiPompeo also would like to thank ESO for the DGDF funding used to support a very productive visit to ESO headquarters in 2009.

{\it Facilities:} \facility{VLT: (FORS2)}

\clearpage
%tables go in here separated by \clearpage statements.

\begin{deluxetable}{lcccccc}
 \tabletypesize{\scriptsize}
 \tablewidth{0pt}
 \tablecaption{The BALQSO sample\label{mainpropstbl}}
 \tablehead{
   \colhead{Source Name} & \colhead{RA} & \colhead{DEC} & \colhead{$z$} & \colhead{$M_B$} & \colhead{Class} & \colhead{Radio source?}
  }
   \startdata
       LBQS 2358+0216              & 00 01 21.6   &  $+$02 33 05   &  1.87  & -27.1\tablenotemark{a} & LoBAL      & no   \\    
       SDSS J0018+0015            & 00 18 24.9   &  $+$00 15 26   &  2.43  & -26.4                                 &  HiBAL      & no   \\
       PKS 0040-005                    & 00 43 23.3   &  $-$00 15 52    &  2.81  & -28.1                                 & HiBAL       & yes  \\
       $[$D87$]$ UJ3682P-057  & 00 49 38.1   &  $-$30 39 51    &  2.36  & -27.2\tablenotemark{b} & HiBAL       & no    \\
       FBQS J0105-0033              & 01 05 40.7   &  $-$00 33 14    &  1.18  & -25.5                                 & HiBAL?     & yes  \\ 
       LBQS 0109-0128                & 01 12 27.5   &  $-$01 12 20    &  1.76  & -27.3                                 & HiBAL       & no    \\
       SDSS J0115-0033              & 01 15 39.8   &  $-$00 33 27    &  1.59  & -27.3                                 & HiBAL       & no    \\
       FBQS J0127+0114             & 01 27 02.5   &  $+$01 14 12   &  1.16  & -25.7                                 & LoBAL?    & yes   \\
       $[$HB89$]$ 0134-376       & 01 36 52.5   &  $-$37 25 04    &  2.52  & -29.1\tablenotemark{c} & HiBAL       & no     \\
       SDSS J0250+0035             & 02 50 42.4   &  $+$00 35 38   &  2.38  & -27.1                                 & LoBAL      & no     \\ 
       SDSS J0300+0048             & 03 00 00.5   &  $+$00 48 33   &  0.89  & -24.4                                 & FeLoBAL & no     \\
       SDSS J0318-0600              & 03 18 56.6   &  $-$06 00 36    &  1.97  & -26.6                                 & FeLoBAL & no      \\
       $[$HB89$]$ 0335-336       & 03 37 21.6   &  $-$33 29 11    &  2.26  & -27.5\tablenotemark{c} & HiBAL       & no      \\ 
       SDSS J0338+0056            & 03 38 10.8   &  $+$00 56 18   &  1.63  & -25.9                                 & LoBAL?    & no      \\
       SDSS J1337-0246             & 13 37 01.3   &  $-$02 46 31    &  3.06  & -28.0                                 & HiBAL       & yes    \\
       SDSS J1516-0056             & 15 16 30.3   &  $-$00 56 27    &  1.92  & -26.7                                 & HiBAL       & yes    \\
       $[$WHO91$]$ 2043-347   & 20 46 44.9   &  $-$34 33 35    &  3.35  & -27.3\tablenotemark{b} & HiBAL       & no      \\
       LBQS 2111-4335                & 21 15 06.7   &  $-$43 23 10    &  1.71  & -28.9\tablenotemark{a} & HiBAL       & no      \\
       FBQS J2204+0031             & 22 04 45.3   &  $+$00 31 42   &  1.35  & -26.2                                 & HiBAL?     & yes    \\
       SDSS J2215-0045              & 22 15 11.8   &  $-$00 45 48    &  1.48  & -27.8                                 & LoBAL       & no      \\
       $[$HB89$]$ 2240-370       & 22 43 46.0   &  $-$36 47 03    &  1.83  & -28.6\tablenotemark{c} & LoBAL        & no      \\
       SDSS J2352+0105             & 23 52 38.0   &  $+$01 05 53   &  2.16  & -28.3                                 & HiBAL        & no      \\ 
       LBQS 2350-0045A              & 23 52 53.4   &  $-$00 28 50    &  1.62  & -26.9                                 & HiBAL        & no      \\
       \enddata
 \tablecomments{Absolute B magnitudes are calculated using SDSS g-band magnitudes unless noted otherwise.  BAL types in the sixth column followed by a ``?" are uncertain- see \S 2 for details. The final column indicates if an object is a known radio source identified in the FIRST survey.  More radio information for these objects is found in table~\ref{radiopropstbl}.}
 \tablenotetext{a}{Based on B band magnitudes from LBQS (Hewett et al. 1995).}
 \tablenotetext{b}{Based on V and B-V measurements from Veron-Cetty \& Veron (2001)}
 \tablenotetext{c}{Based on measurements from the total light spectra obtained during spectropolarimetric observations.}
\end{deluxetable}

\clearpage

\begin{deluxetable}{cccccc}
 \tabletypesize{\scriptsize}
 \tablewidth{0pt}
 \tablecaption{Properties of radio sources\label{radiopropstbl}}
 \tablehead{
  \colhead{Object} & \colhead{$S_p$} & \colhead{$S_i$} & \colhead{log$L_r$} & \colhead{log$R^*$} & \colhead{$\alpha_r$}
  }
  \startdata                                       
  FBQS 0105$-$0033\tablenotemark{a}  & 4.20$\pm$0.11                                      & 4.0$\pm$0.5                                      & 32.5                                        & 1.2                                          & \nodata                         \\
  FBQS 0127$+$0114\tablenotemark{a} & 1.03$\pm$0.14                                      & 1.2$\pm$0.14\tablenotemark{b}    & 32.2                                        & 1.1                                          & \nodata                         \\
  FBQS 2204$+$0031\tablenotemark{a} & 2.40$\pm$0.11                                      & 3.6$\pm$0.4                                      & 32.6                                        & 0.9                                          & \nodata                          \\
  SDSS 0043$-$0015                                  & 130.68$\pm$0.10\tablenotemark{c} & 167.4$\pm$5.9                                  & 34.3                                        & 2.9                                          & -1.2\tablenotemark{d} \\
  SDSS 1337$-$0246                                  & 43.32$\pm$0.14                                   & 45.1$\pm$1.4                                    & 33.5                                        & 2.1                                          & \nodata                           \\
  SDSS 1516$-$0056                                  & 24.58$\pm$0.14                                   & 25.8$\pm$0.9                                    & 33.2                                        & 1.8                                          & \nodata                           \\
 \enddata
 \tablecomments{$S_p$ is the 1.4 GHz FIRST peak flux (mJy beam$^{-1}$) and $S_i$ is the 1.4 GHz integrated NVSS flux (mJy).  $L_r$ and $R^{*}$ are the K-corrected 5 GHz radio luminosity and radio-loudness parameter, respectively, as defined in Stocke et al. (1992).  $\alpha_r$ is the radio spectral index.  See text for a full discussion of the parameters.}
 \tablenotetext{a}{These sources are not classified as BALs with certainty- see \S 2 and Hall et al. (2002).}
 \tablenotetext{b}{The radio flux form this source is below the limit of NVSS; this value is the FIRST integrated flux.}
 \tablenotetext{c}{The core of this source is too faint for FIRST detection; this value is the peak flux in the lobes.}
 \tablenotetext{d}{As reported in Brotherton et al. (2006).}
\end{deluxetable}

\clearpage

\begin{deluxetable}{ccccccc}
 \tabletypesize{\scriptsize}
 \tablewidth{0pt}
 \tablecaption{Continuum and white light polarization properties of the sample.\label{contpoltbl}}
 \tablehead{
   \colhead{Object} & \colhead{White P(\%)} & \colhead{White PA($^{\circ}$)} & \colhead{Cont. P(\%)} & \colhead{Cont. PA($^{\circ}$)} & \colhead{Cont. $\lambda$(\AA)} & \colhead{Max. ISP (\%)}
   }
  \startdata
0001+0233	& 0.63$\pm$0.07 & 43$\pm$3   & 0.94$\pm$0.19          & 37$\pm$6	     & 4570-5100 & 0.23 \\
0018+0015	& 6.54$\pm$0.09 & 53$\pm$1   & 7.78$\pm$0.12          & 54$\pm$1      & 5450-6200 & 0.24 \\
0043$-$0015	& 0.61$\pm$0.08 & 128$\pm$4 & 0.51$\pm$0.19          & 128$\pm$11 & 6100-6800 & 0.15 \\
0049$-$3039	& 2.41$\pm$0.09 & 48$\pm$1   & 2.83$\pm$0.18          & 55$\pm$2      & 5310-6115 & 0.19 \\
0105$-$0033	& 0.46$\pm$0.07 & 172$\pm$4 & 0.31$\pm$0.12          & 153$\pm$13 & 4200-5650 & 0.32 \\
0112$-$0112	& 1.66$\pm$0.06 & 68$\pm$1   & 1.78$\pm$0.14          & 66$\pm$2      & 5500-6200 & 0.63 \\
0115$-$0033	& 0.78$\pm$0.05 & 157$\pm$2 & 0.94$\pm$0.10          & 156$\pm$3   & 5600-6800 & 0.27 \\
0127+0114	& 7.10$\pm$0.06 & 103$\pm$1 & 8.07$\pm$0.12          & 103$\pm$1   & 4240-5650 & 0.25 \\
0136$-$3725	& 0.32$\pm$0.05 & 85$\pm$4   & 0.24$\pm$0.11          & 88$\pm$15    & 5600-6135 & 0.14 \\
0250+0035	& 4.03$\pm$0.09 & 68$\pm$1   & 4.30$\pm$0.20          & 66$\pm$1      & 5445-6060 & 0.55 \\
0300+0048	& 1.77$\pm$0.02 & 10$\pm$1   & 1.94$\pm$0.07          & 10$\pm$1      & 6360-6600 & 0.80 \\
0318$-$0600	& 2.99$\pm$0.05 & 164$\pm$1 & 3.35$\pm$0.20          & 172$\pm$2   & 5050-5320 & 0.44 \\
0337$-$3329	& 0.67$\pm$0.09 & 30$\pm$4   & 0.88$\pm$0.22          & 25$\pm$7      & 5110-5780 & 0.81 \\
0338+0056	& 0.74$\pm$0.09 & 42$\pm$3   & 0.71$\pm$0.35          & 8$\pm$16      & 4400-4900 & 0.96 \\
1337$-$0246    & 1.35$\pm$0.04 & 58$\pm$1   & 1.86$\pm$0.11          & 75$\pm$2      & 5710-6125 & 0.47 \\
1516$-$0056   & 2.37$\pm$0.04  & 54$\pm$1   & 2.08$\pm$0.10         & 58$\pm$2      & 4960-5400 & 0.63 \\
2046$-$3433	& 1.08$\pm$0.15 & 163$\pm$4 & 0.64$\pm$0.34          & 160$\pm$17 & 6900-7570 & 0.56 \\
2115$-$4323	& 0.85$\pm$0.04 & 97$\pm$1   & 0.88$\pm$0.06          & 99$\pm$2      & 5410-6820 & 0.33 \\
2204+0031	& 0.18$\pm$0.06 & 7$\pm$9     & 0.22$\pm$0.13          & 145$\pm$18  & 4500-5500 & 0.50 \\
2215$-$0045	& 0.46$\pm$0.03 & 82$\pm$2   & 0.30$\pm$0.08          & 82$\pm$6      & 6015-6500 & 0.95 \\
2243$-$3647	& 2.42$\pm$0.03 & 26$\pm$1   & 2.54$\pm$0.06          & 25$\pm$1      & 4500-5150 & 0.12 \\
2352+0105	& 1.46$\pm$0.03 & 23$\pm$1   & 1.62$\pm$0.05          & 23$\pm$1      & 4960-5670 & 0.23 \\
2352$-$0028	& 0.60$\pm$0.07 & 86$\pm$3   & 0.87$\pm$0.27          & 84$\pm$ 9     & 4130-4460 & 0.26 \\
  \enddata
 \tablecomments{White light polarization measurements are averaged over the whole spectrum, from 4000-8000 \AA.  Continuum polarization measurements are averaged over the (observed) wavelength region listed in column 6.  Uncertainties on polarization are not necessarily symmetric; values reported are an average of upper and lower uncertainties.  Maximum interstellar polarizations along the line of sight using the Serkowsky Law are listed in column 7.}
\end{deluxetable}

\clearpage

\begin{deluxetable}{ccccccccccc}
 \tabletypesize{\scriptsize}
 \rotate
 \tablewidth{0pt}
 \tablecaption{Polarization properties of emission and absorption features.\label{linepoltbl}}
 \tablehead{
    \colhead{Object} & \colhead{$p_e$ (\ion{C}{4})} & \colhead{$p_e$ (\ion{C}{3}])} & \colhead{$p_e$ (\ion{Mg}{2})} & \colhead{$p_a$ (\ion{C}{4})} & \colhead{$p_a$ (\ion{Mg}{2})} & \colhead{$\frac{p_e}{p_c}$ (\ion{C}{4})} & \colhead{$\frac{p_e}{p_c}$ (\ion{C}{3}])} & \colhead{$\frac{p_e}{p_c}$ (\ion{Mg}{2})} & \colhead{$\frac{p_a}{p_c}$ (\ion{C}{4})} & \colhead{$\frac{p_a}{p_c}$ (\ion{Mg}{2})}
  }
   \startdata
   0001+0233	 &  $<$0.67 & 0.45         & 0.98         & 1.32        & \nodata & $<$0.63 & 0.39        & 0.96         & 1.25        & \nodata  \\
   0018+0015	 &  4.03        & 5.41         & \nodata   & 8.39        & \nodata & 0.51        & 0.68         & \nodata   & 1.06        &  \nodata  \\
   0043$-$0015	 &  0.61        & \nodata   & \nodata   & 1.13        & \nodata & 0.85        & \nodata   & \nodata    & 1.57        &  \nodata  \\
   0049$-$3039	 &  1.32        & 2.35         & \nodata   & 7.12        & \nodata & 0.49        & 0.88        & \nodata    & 2.66        &  \nodata   \\ 
   0105$-$0033	 &  \nodata  & \nodata    & \nodata   & \nodata  & \nodata & \nodata  & \nodata   & \nodata   & \nodata   &  \nodata  \\          
   0112$-$0112	 &  1.18        & 1.25         & 0.57         & 3.25        & \nodata & 0.50        & 0.71        & 0.29         & 1.38        &  \nodata  \\ 
   0115$-$0033	 &  0.52        & 0.54         & \nodata   & $<$0.57 & \nodata & 1.37        & 0.56        & \nodata    & $<$1.50 &   \nodata \\ 
   0127+0114	 &  \nodata  & 6.41         & 7.16         & \nodata   & 7.69      & \nodata   & 0.76       & 1.05          & \nodata  &   1.13       \\ 
   0136$-$3725	 &  $<$0.28 & 0.55         & \nodata   & $<$0.45 & \nodata & $<$0.78 & 1.62        & \nodata    & $<$1.25 &  \nodata  \\ 
   0250+0035	 &  2.20        & 3.91         & \nodata   & 3.31        & \nodata & 0.60        & 0.94        & \nodata    & 0.91        & \nodata   \\ 
   0300+0048	 & \nodata   & \nodata    & \nodata   & \nodata  & \nodata & \nodata   & \nodata  & \nodata    & \nodata  & \nodata  \\
   0318$-$0600	 & \nodata   & 2.85         & 3.41         & 2.29        & 3.12      & \nodata    & 0.96       & 1.11          & 0.62        & 1.02         \\ 
   0337$-$3329	 &  $<$0.82 & 0.73         & \nodata   & $<$0.44 & \nodata & $<$0.99 & 0.55        & \nodata    & $<$0.53 & \nodata   \\
   0338+0056	 &  2.79        & 1.25         & 2.11         & \nodata  & \nodata & 1.25        & 1.28        & 1.80          & \nodata  & \nodata   \\
   1337$-$0246 & \nodata    & \nodata   & \nodata   & 3.55        & \nodata & \nodata  & \nodata   & \nodata    & 1.92        & \nodata   \\
   1516$-$0056 & 2.69          & 3.93        & \nodata   & 2.68        & \nodata & 1.23        & 1.45        & \nodata    & 1.23         & \nodata   \\
   2046$-$3433	 &  1.79         & $<$1.77 & \nodata   &$<$1.96  & \nodata & 2.59        & $<$1.08 & \nodata    & $<$2.84 & \nodata    \\
   2115$-$4323	 &  0.91         & 0.67        & 0.67        &1.77          & \nodata & 0.90        & 0.66        & 0.84         & 1.74         &  \nodata   \\
   2204+0031	 & \nodata    & \nodata   & \nodata  & \nodata   & \nodata  & \nodata & \nodata   &\nodata    & \nodata    & \nodata    \\
   2215$-$0045	 &  0.41         & 0.61        & $<$0.40 & $<$0.74 & 0.32      & 1.78         & 1.49         & $<$1.29 & $<$3.22   & 1.03          \\
   2243$-$3647	 &  1.75         & 2.06        & 1.52        & 3.09        & 1.87       & 0.71        & 0.78         & 0.77        & 1.25          & 0.95          \\
   2352+0105	 &  \nodata   & 1.61        & \nodata   & 1.14        & \nodata & \nodata  & 1.09         & \nodata   & 0.66          & \nodata    \\
   2352$-$0028	 &  1.12         & 0.73        & 0.51        & 4.02        & \nodata & 1.29        & 1.49         & 1.19         & 4.62          & \nodata    \\
      \enddata
  \tablecomments{Typical uncertainties on $p_e$ and $p_a$ are approximately 0.5\%.  See text for discussion of all parameters.  Values preceded by ``$<$" are upper limits.}
\end{deluxetable}

\clearpage

\begin{deluxetable}{ccccc}
 \tabletypesize{\scriptsize}
 \tablewidth{0pt}
 \tablecaption{Correlation analysis of polarization properties.\label{corrtbl}}
 \tablehead{
   \colhead{Property 1} & \colhead{Property 2} & \colhead{$r_s$} & \colhead{$P_{r_s}$} & \colhead{n}
   }
  \startdata                
  Cont. $p$                          & $M_B$     	                   &  0.547    & \textbf{0.015} & 19  \\
  Cont. $p$                          & $p_e/p_c$(\ion{C}{4})     &  -0.727   & \textbf{0.007} & 12  \\
  Cont. $p$                          & $p_e/p_c$(\ion{C}{3}])    &  -0.209   & 0.453              & 15  \\
  Cont. $p$                          & $p_e/p_c$(\ion{Mg}{2})  &  \nodata  & \nodata          & 6     \\
  Cont. $p$                          & $p_a/p_c$(\ion{C}{4})     &  -0.514   & 0.072              & 13   \\
  Cont. $p$                          & $p_a/p_c$(\ion{Mg}{2})  &   \nodata & \nodata          & 3      \\      
  $p_e/p_c$(\ion{C}{4})    & $p_e/p_c$(\ion{C}{3}])    &   0.216    & 0.575              & 10    \\                     
  $p_e/p_c$(\ion{C}{4})    & $p_e/p_c$(\ion{Mg}{2})  &   \nodata & \nodata           &  4   \\ 
  $p_e/p_c$(\ion{C}{4})    & $p_a/p_c$(\ion{C}{4})     &   0.216    & 0.575              & 9   \\
  $p_e/p_c$(\ion{C}{4})    & $p_a/p_c$(\ion{Mg}{2})  &   \nodata & \nodata           & 2   \\
  $p_e/p_c$(\ion{C}{3}])    & $p_e/p_c$(\ion{Mg}{2}) &   \nodata & \nodata           & 6   \\
  $p_e/p_c$(\ion{C}{3}])    & $p_a/p_c$(\ion{C}{4})    &  -0.154    & 0.649              & 11 \\
  $p_e/p_c$(\ion{C}{3}])    & $p_a/p_c$(\ion{Mg}{2}) &   \nodata & \nodata           &  3   \\
  $p_e/p_c$(\ion{Mg}{2}) & $p_a/p_c$(\ion{C}{4})     &   \nodata & \nodata           &  6   \\
  $p_e/p_c$(\ion{Mg}{2}) & $p_a/p_c$(\ion{Mg}{2})   &   \nodata & \nodata          & 2   \\
  $p_a/p_c$(\ion{C}{4})    & $p_a/p_c$(\ion{Mg}{2})   &  \nodata & \nodata           &  2  \\				       
 \enddata
 \tablecomments{The column labeled $r_s$ gives the Spearman rank correlation coefficient, and $P_{r_s}$ is the corresponding P-value.  Correlations were only looked for when 9 or more data points were available, though all property pairs are listed for completeness.  Values shown in bold type are those that satisfy our (loose) cutoff of $P_{r_s} \le 0.05$.  The number of points considered in each correlation is given in the column labeled $n$.}
\end{deluxetable}

\clearpage

\begin{figure}
 \centering
  \figurenum{2}
   \subfloat[][Fig. 2$a$]{\includegraphics[width=5in]{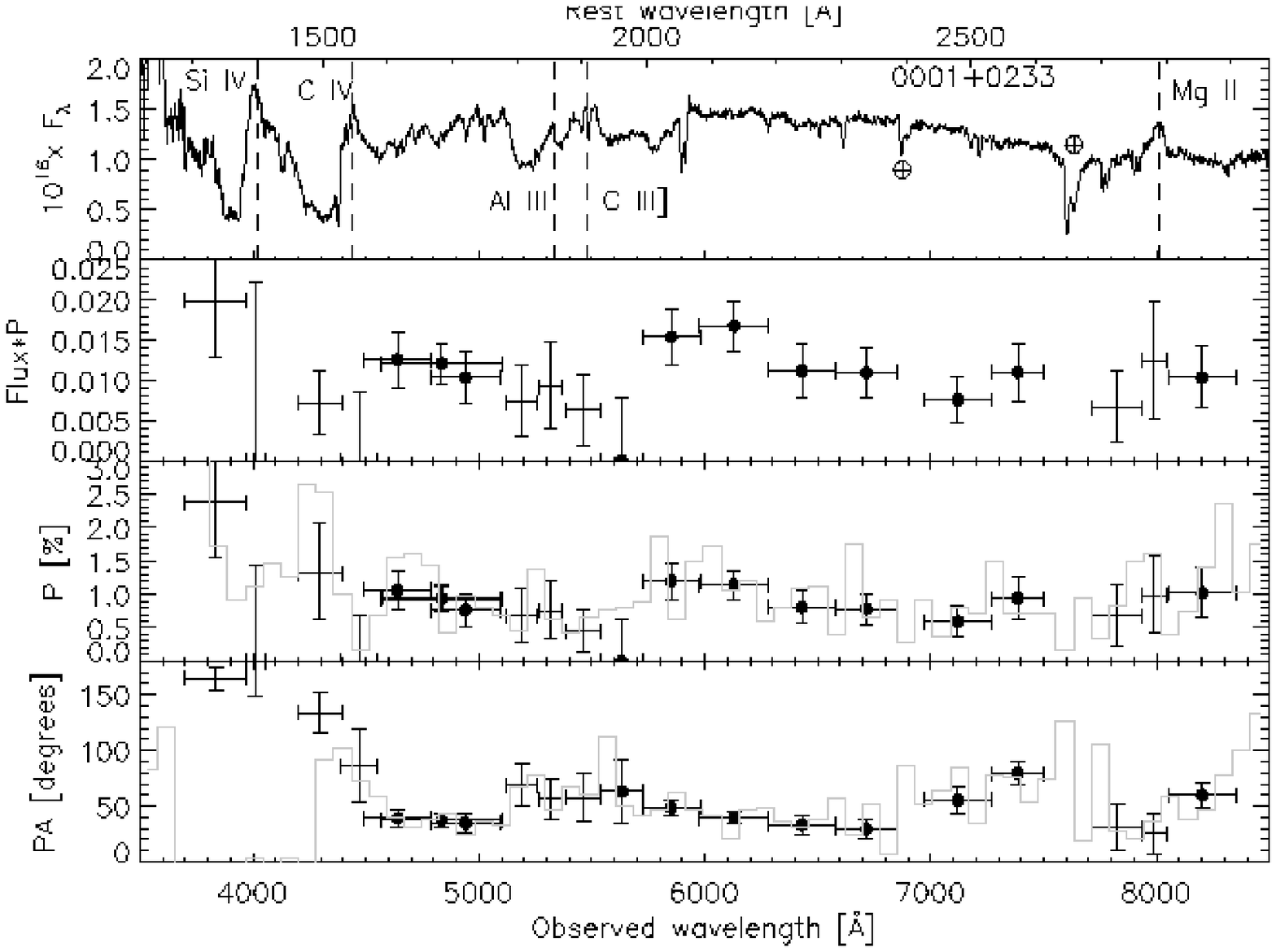}}
  \caption{The following figures show the spectropolarimetry for our whole sample.  The top panel is the total flux spectrum, followed by the polarized flux spectrum ($p$ times $f$), the polarization level, and finally the polarization position angle.  Error bars are 1-$\sigma$ values. The light colored lines in the polarization and position angle plots are smaller equal sized bins (unbiased, with errors omitted) which range from 15-35 \AA\ wide depending on the signal-to-noise of the observations.  Observed frame wavelengths are given on the bottom, and rest frame wavelengths are given along the top.  The name of each object is in the total flux panel.  Prominent lines are marked in the first figure as an aid to the reader; lines marked with $\oplus$ are telluric lines.  Bins plotted in bold in the polarization panel indicate where continuum measurements were made.  Points plotted with a closed circle are continuum measurements, those with just a point are line measurements.\label{polplotsfig}}
\end{figure}

\begin{figure}
 \ContinuedFloat
 \centering
  \subfloat[][Fig. 2$b$]{\includegraphics[width=5in]{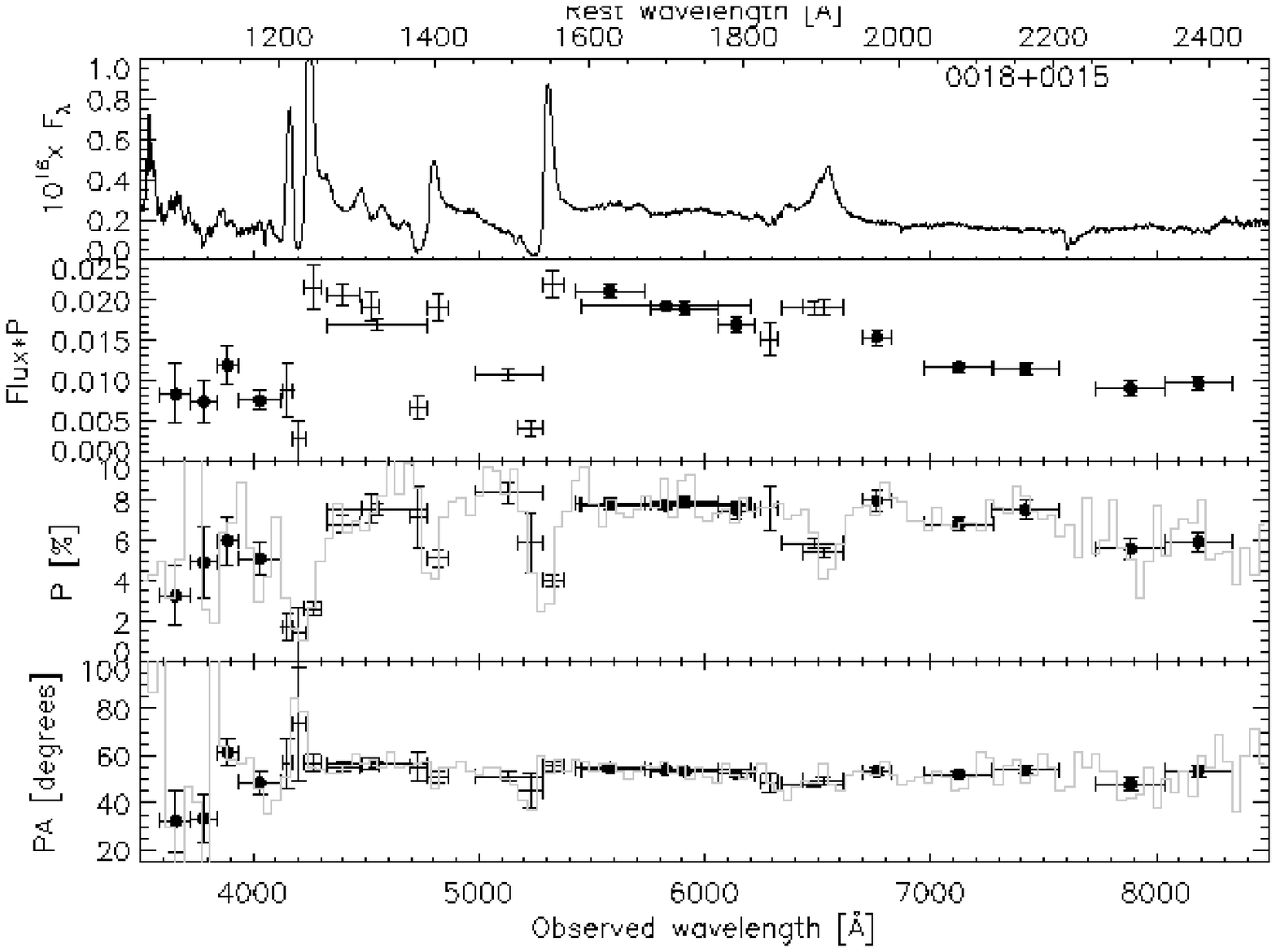}}
\end{figure}

\begin{figure}
 \ContinuedFloat
 \centering
  \subfloat[][Fig. 2$c$]{\includegraphics[width=5in]{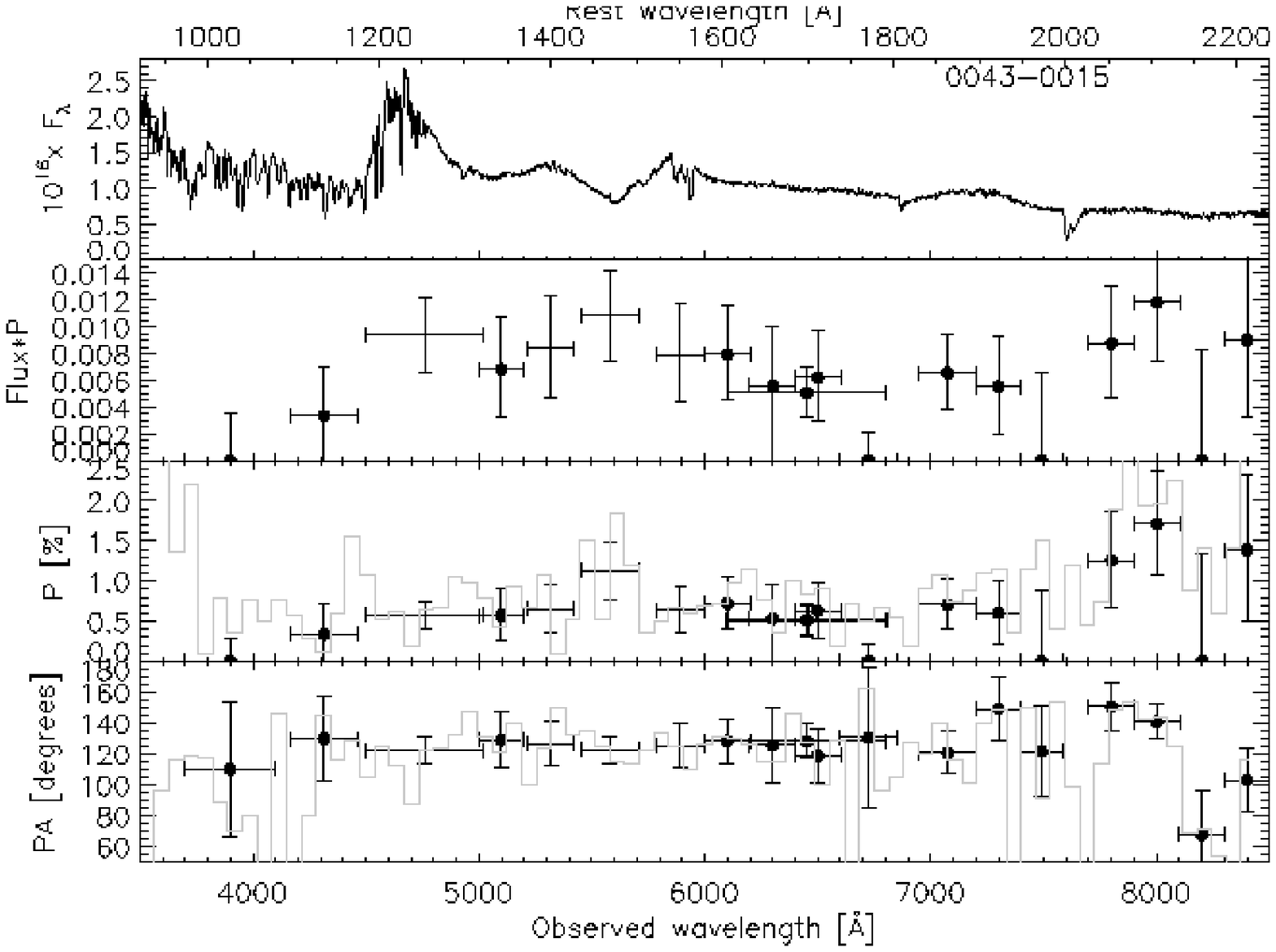}}
\end{figure}

\begin{figure}
 \ContinuedFloat
 \centering
  \subfloat[][Fig. 2$d$]{\includegraphics[width=5in]{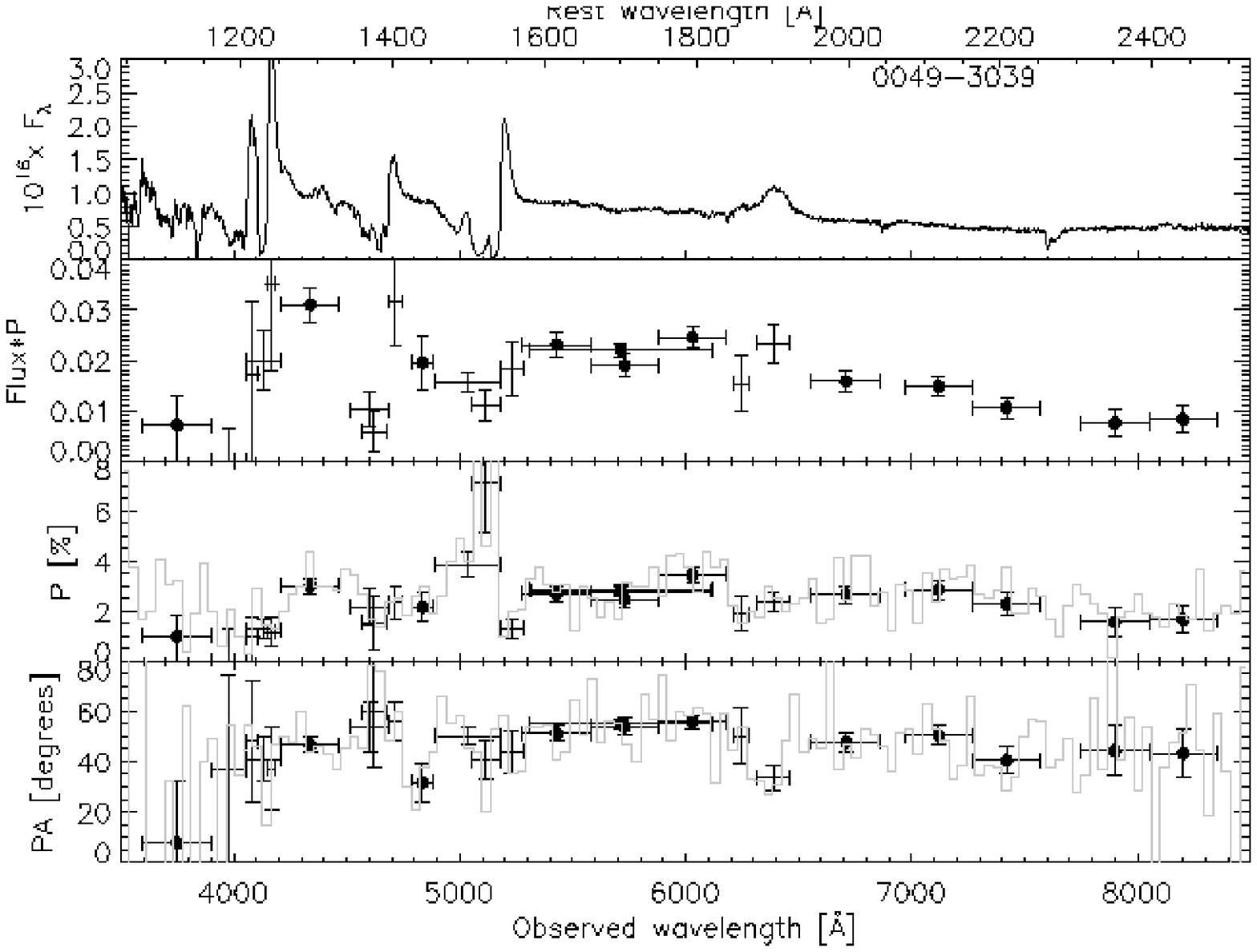}}
\end{figure}

\begin{figure}
 \ContinuedFloat
 \centering
  \subfloat[][Fig. 2$e$]{\includegraphics[width=5in]{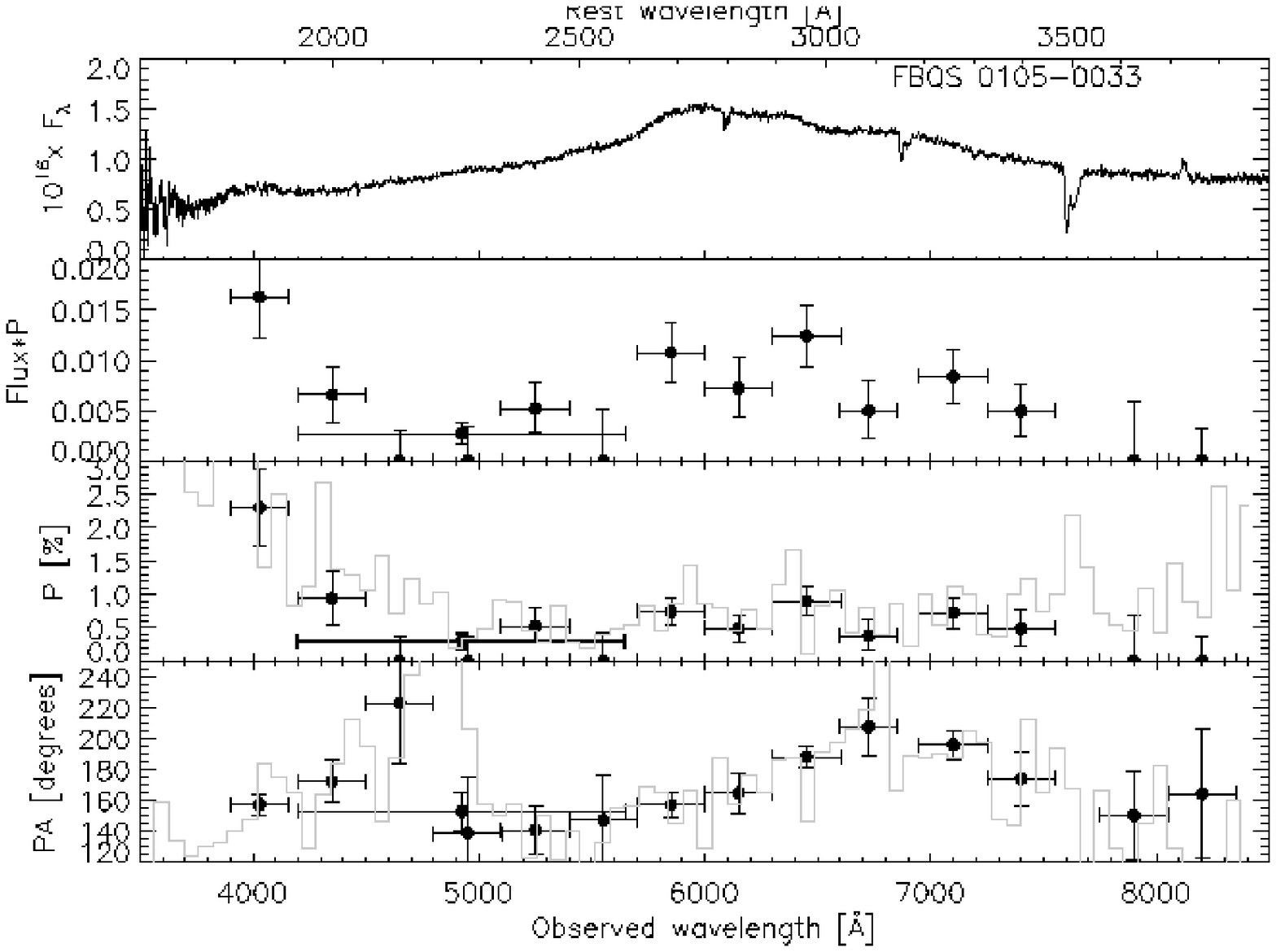}}
\end{figure}

\begin{figure}
 \ContinuedFloat
 \centering
  \subfloat[][Fig. 2$f$]{\includegraphics[width=5in]{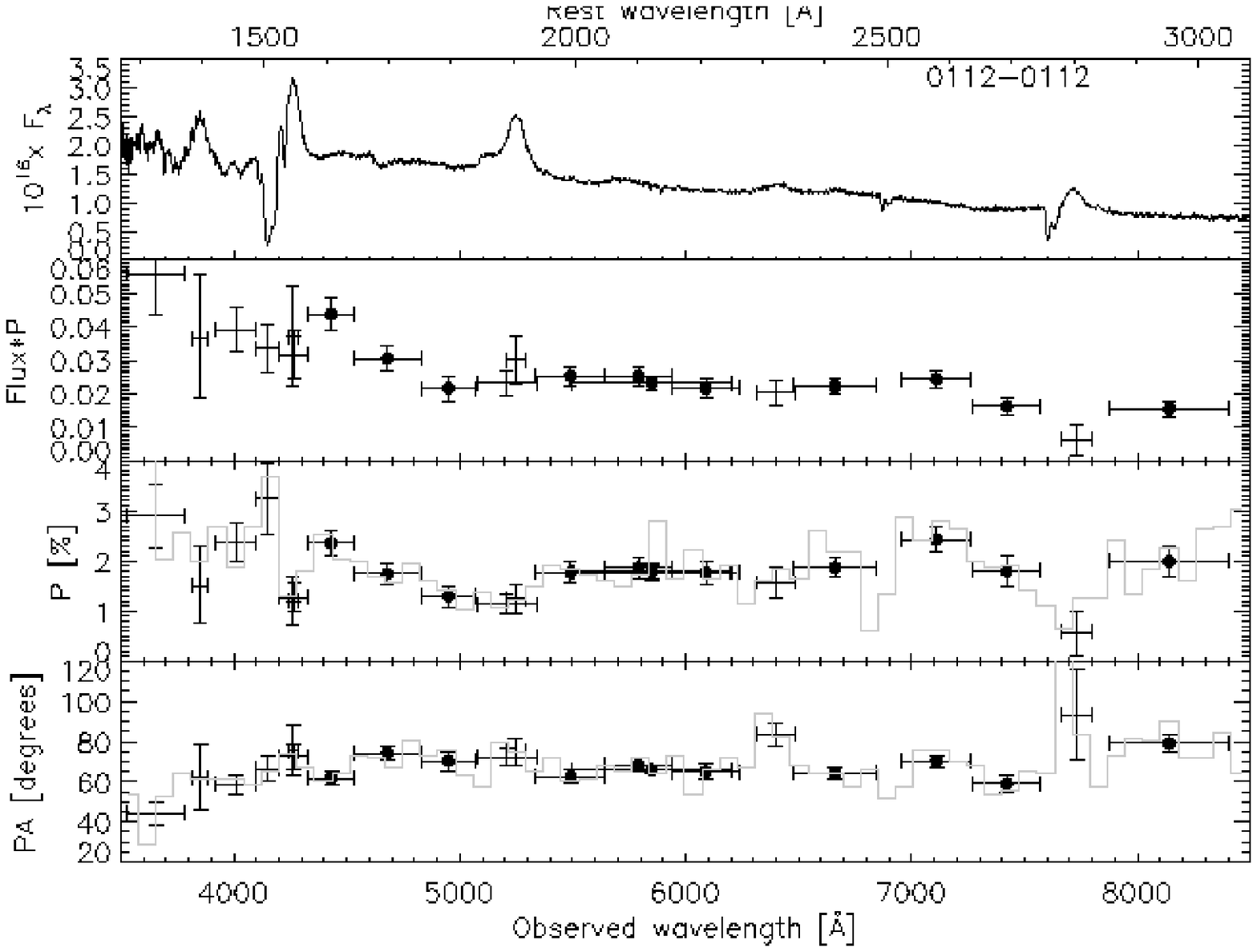}}
\end{figure}

\begin{figure}
 \ContinuedFloat
 \centering
  \subfloat[][Fig. 2$g$]{\includegraphics[width=5in]{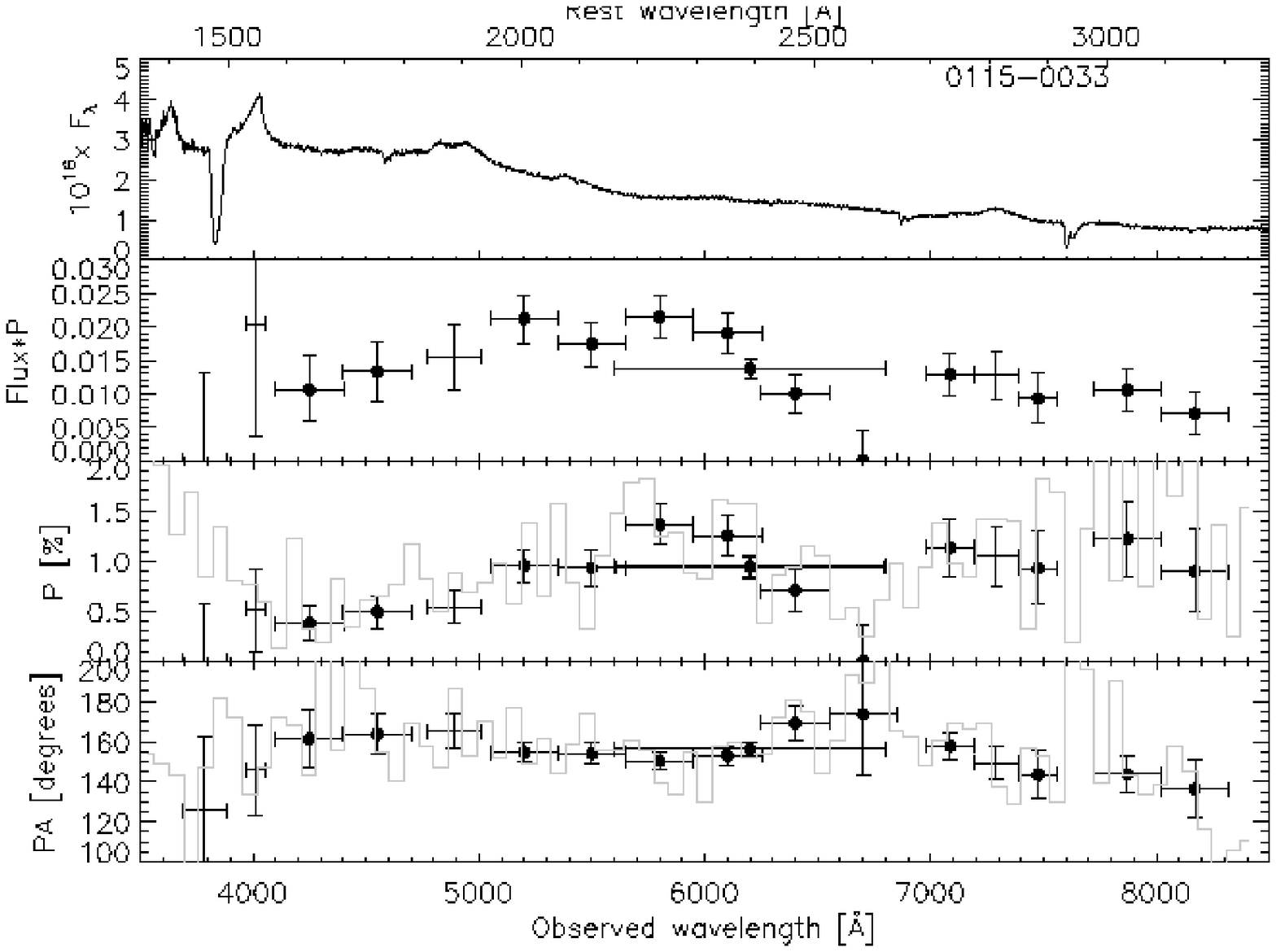}}
\end{figure}

\begin{figure}
 \ContinuedFloat
 \centering
  \subfloat[][Fig. 2$h$]{\includegraphics[width=5in]{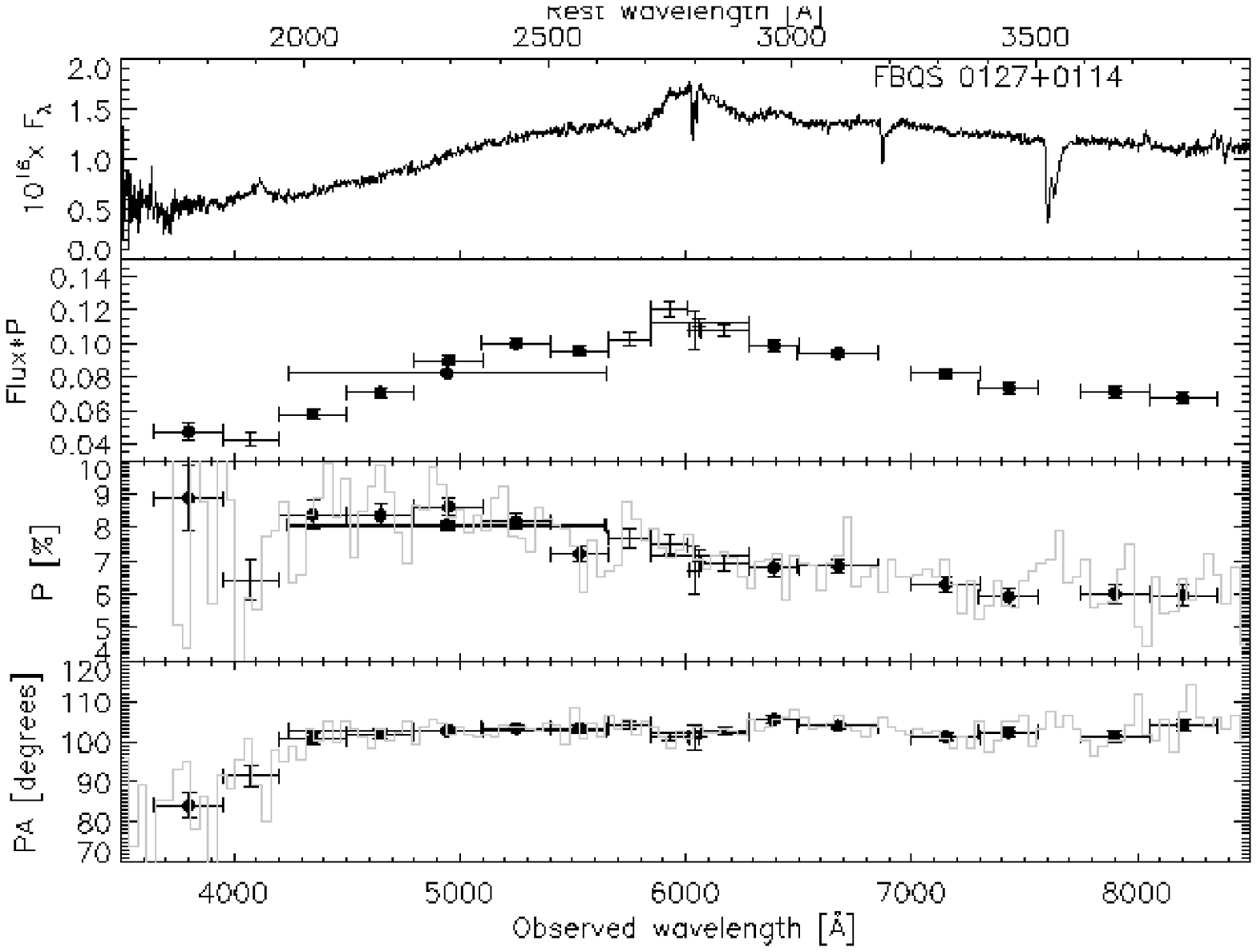}}
\end{figure}

\begin{figure}
 \ContinuedFloat
 \centering
  \subfloat[][Fig. 2$i$]{\includegraphics[width=5in]{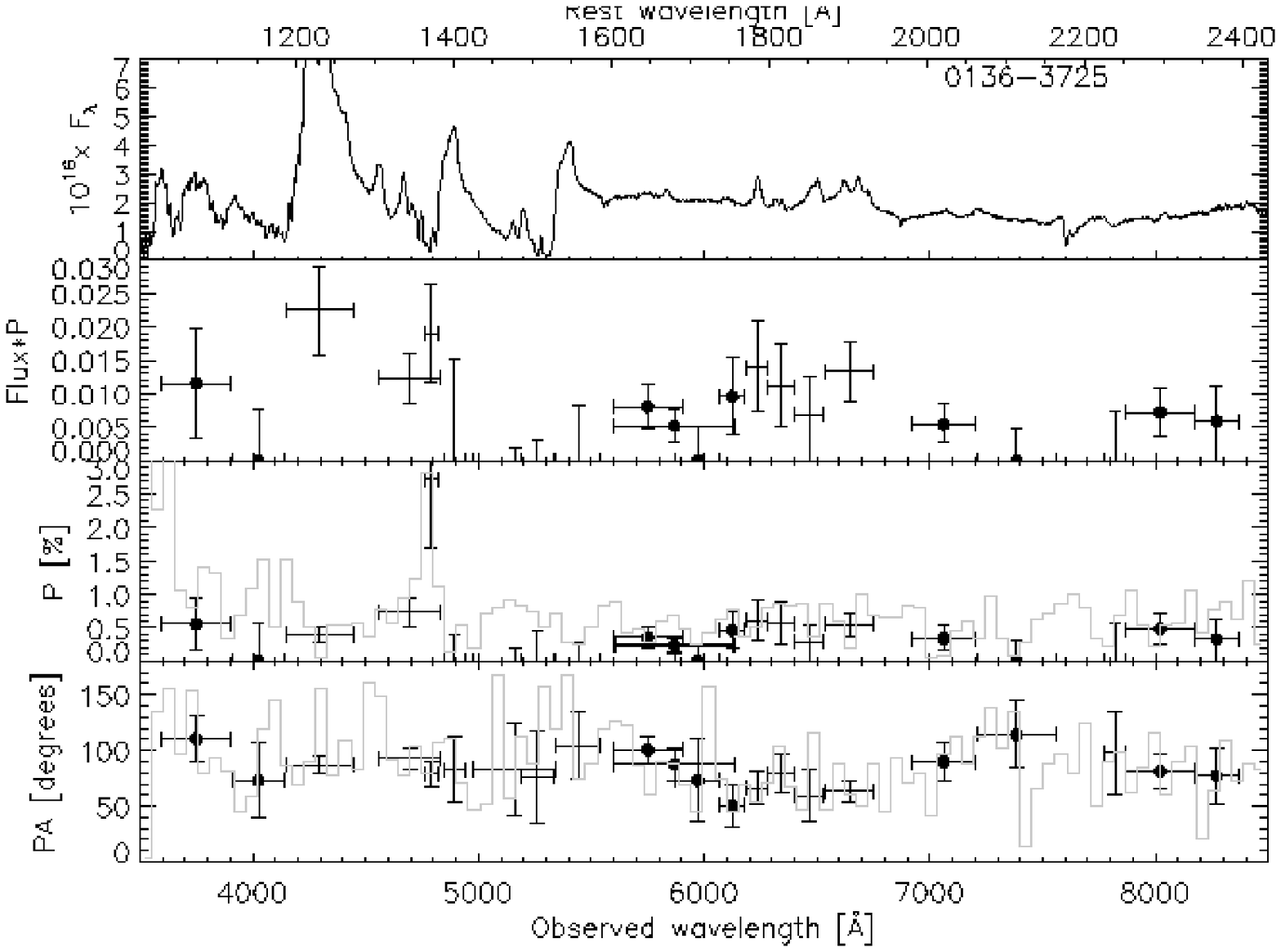}}
\end{figure}

\begin{figure}
 \ContinuedFloat
 \centering
  \subfloat[][Fig. 2$j$]{\includegraphics[width=5in]{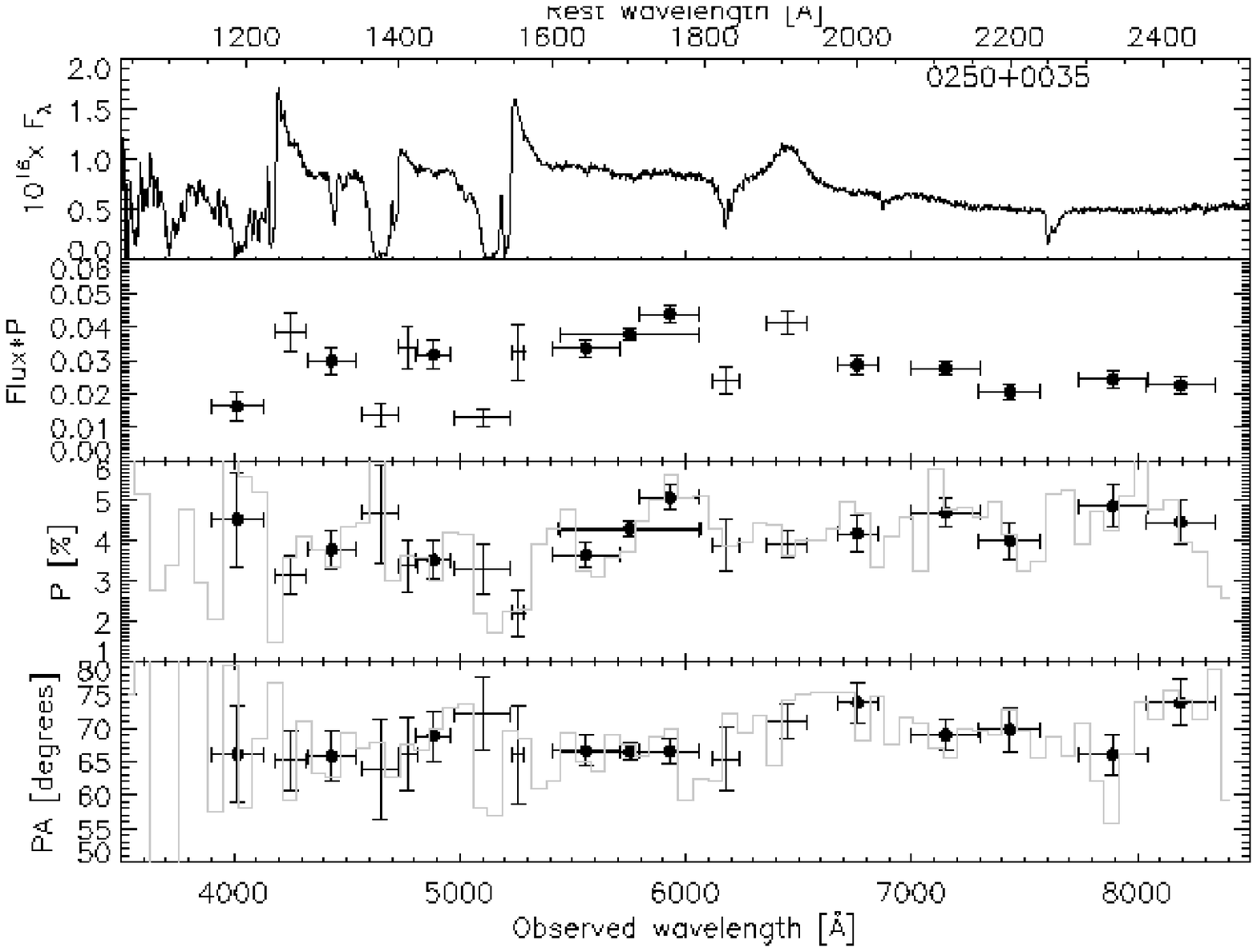}}
\end{figure}

\begin{figure}
 \ContinuedFloat
 \centering
  \subfloat[][Fig. 2$k$]{\includegraphics[width=5in]{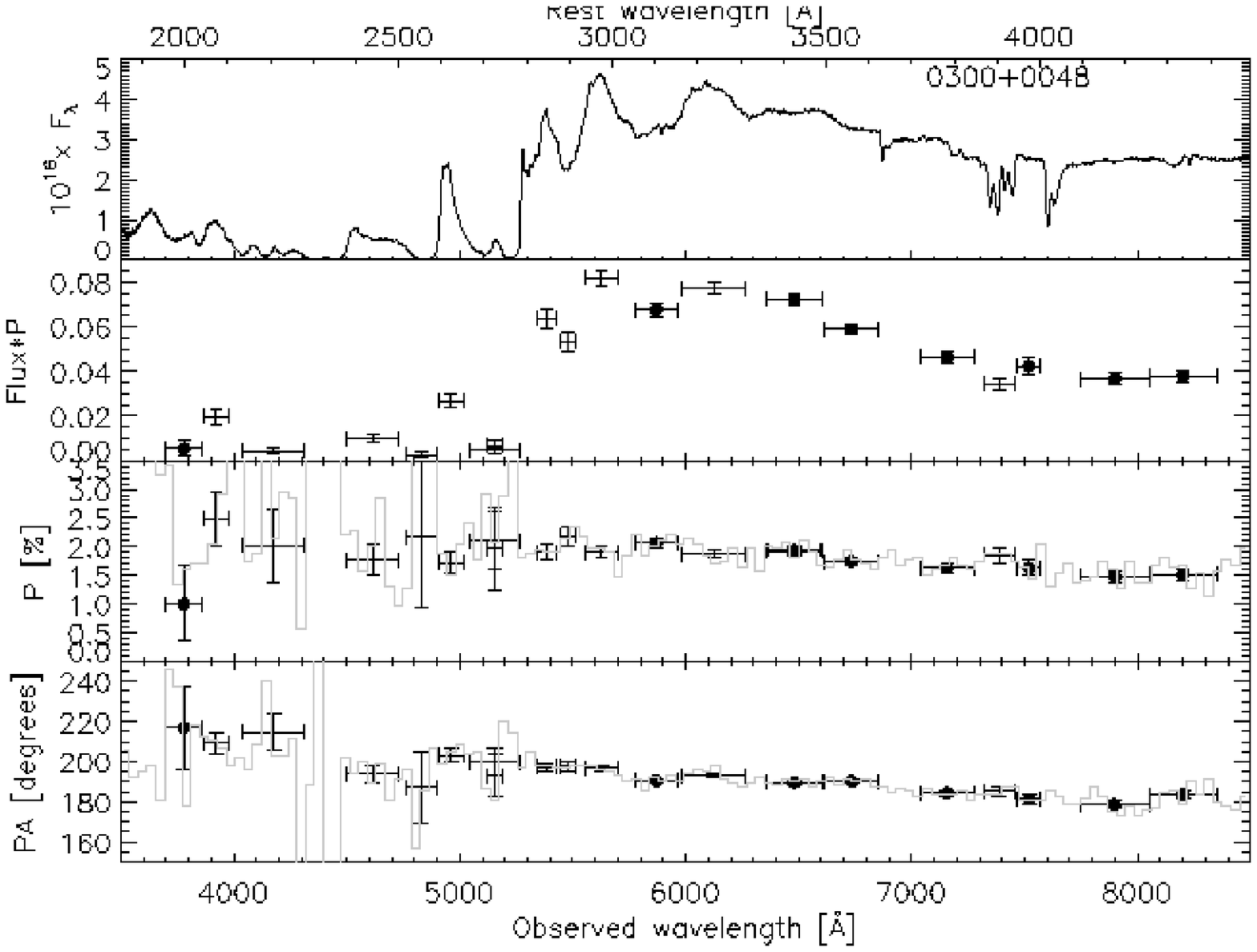}}
\end{figure}

\begin{figure}
 \ContinuedFloat
 \centering
  \subfloat[][Fig. 2$l$]{\includegraphics[width=5in]{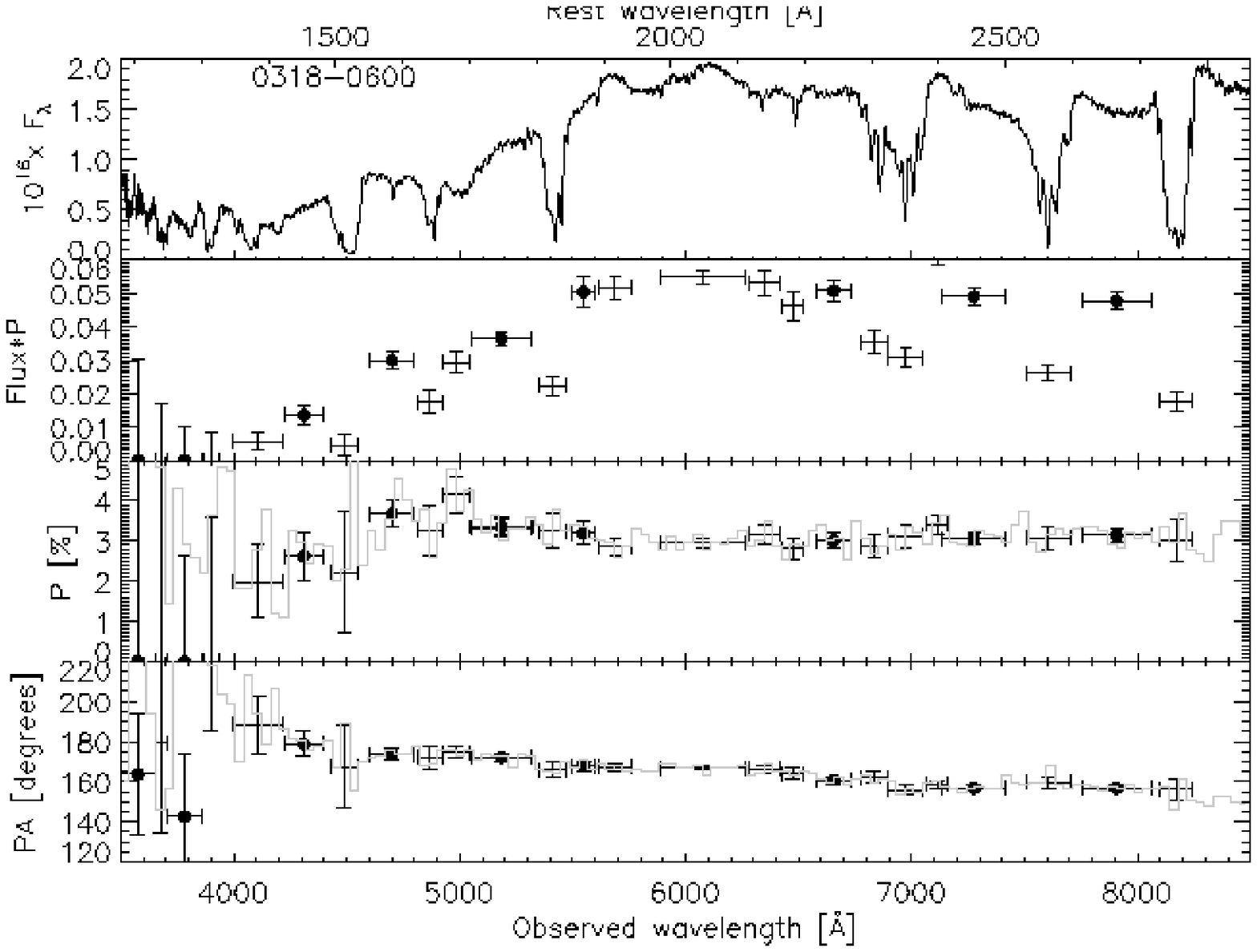}}
\end{figure}

\begin{figure}
 \ContinuedFloat
 \centering
  \subfloat[][Fig. 2$m$]{\includegraphics[width=5in]{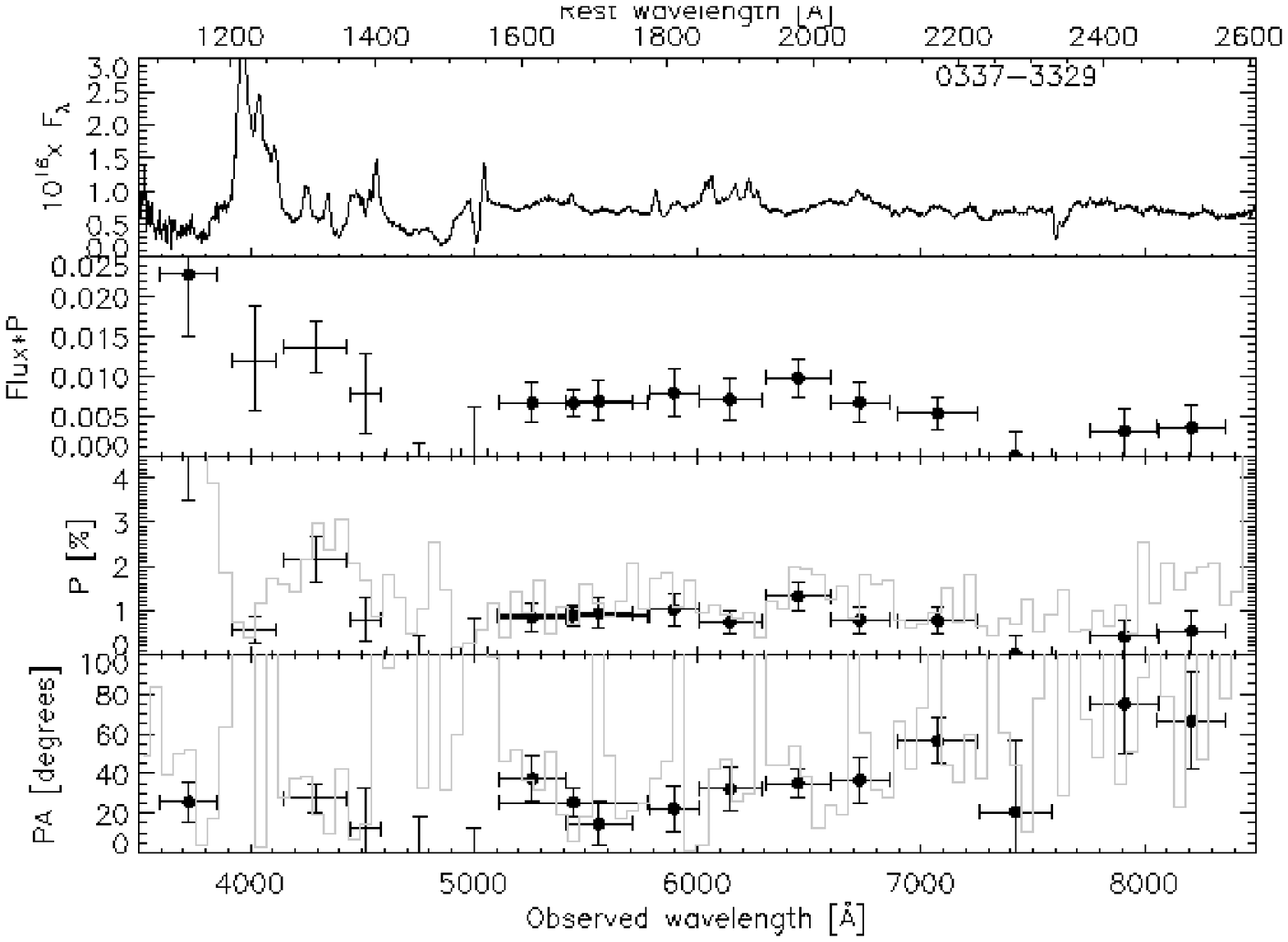}}
\end{figure}

\begin{figure}
 \ContinuedFloat
 \centering
  \subfloat[][Fig. 2$n$]{\includegraphics[width=5in]{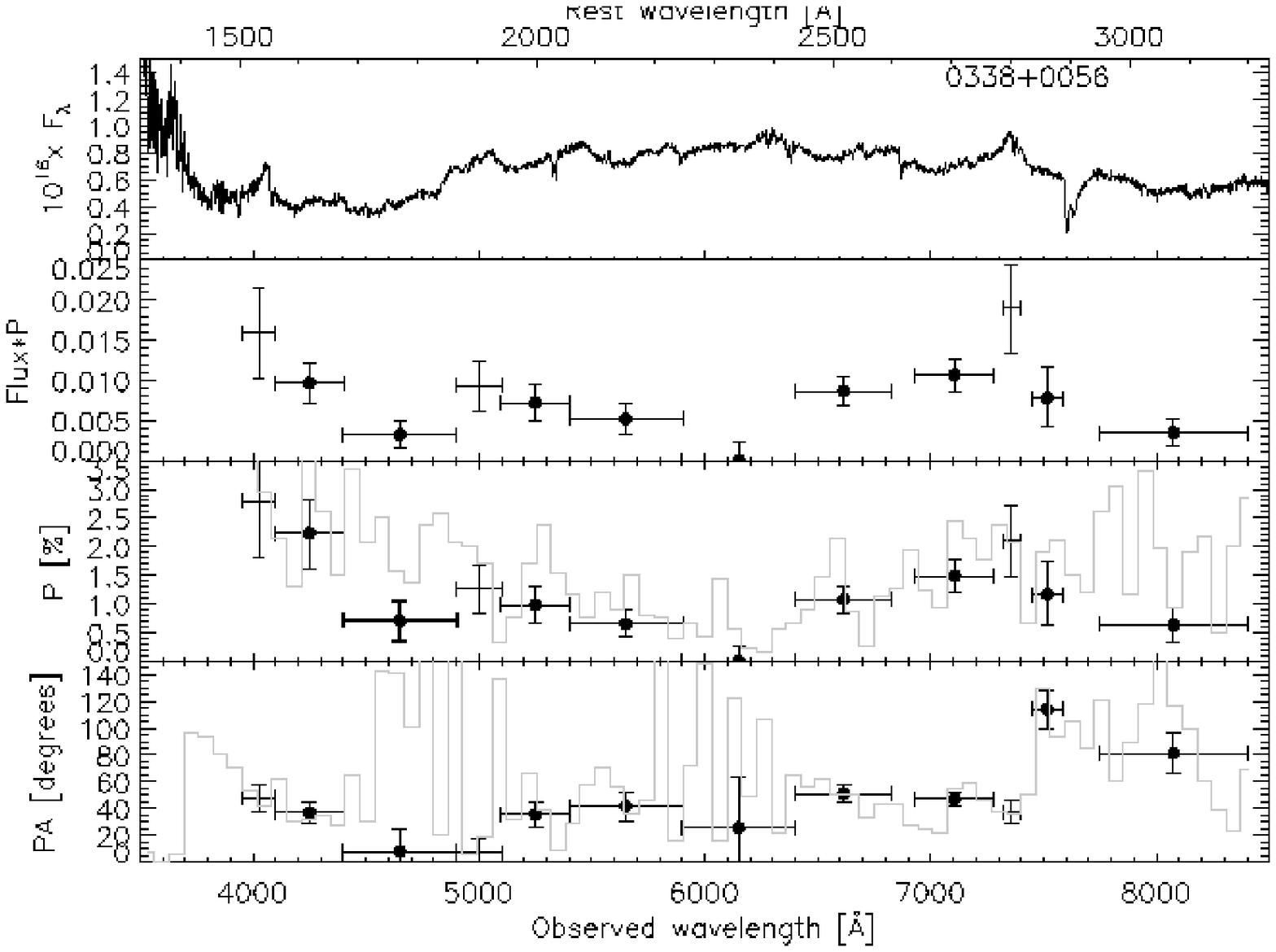}}
\end{figure}

\begin{figure}
 \ContinuedFloat
 \centering
  \subfloat[][Fig. 2$o$]{\includegraphics[width=5in]{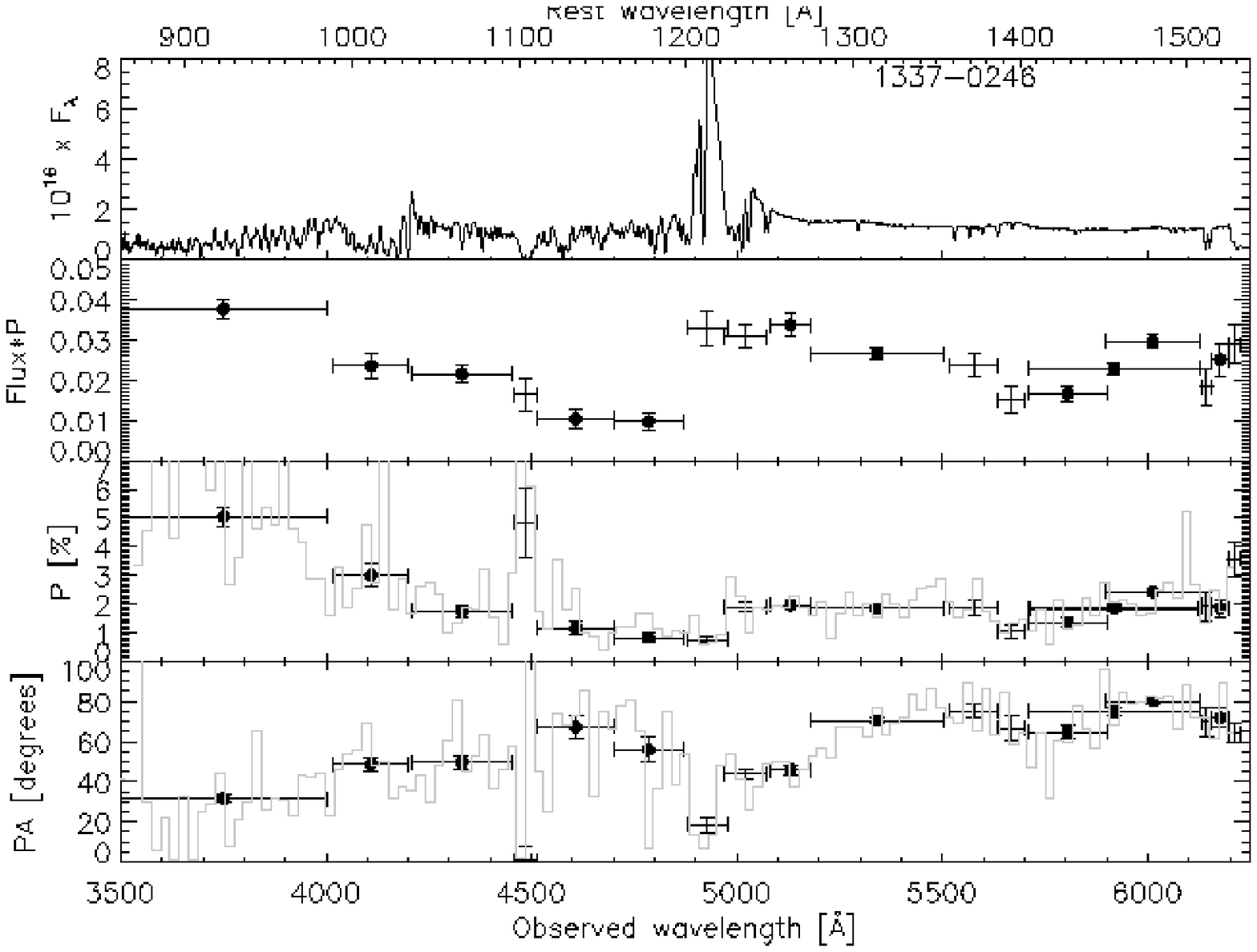}}
\end{figure}

\begin{figure}
 \ContinuedFloat
 \centering
  \subfloat[][Fig. 2$p$]{\includegraphics[width=5in]{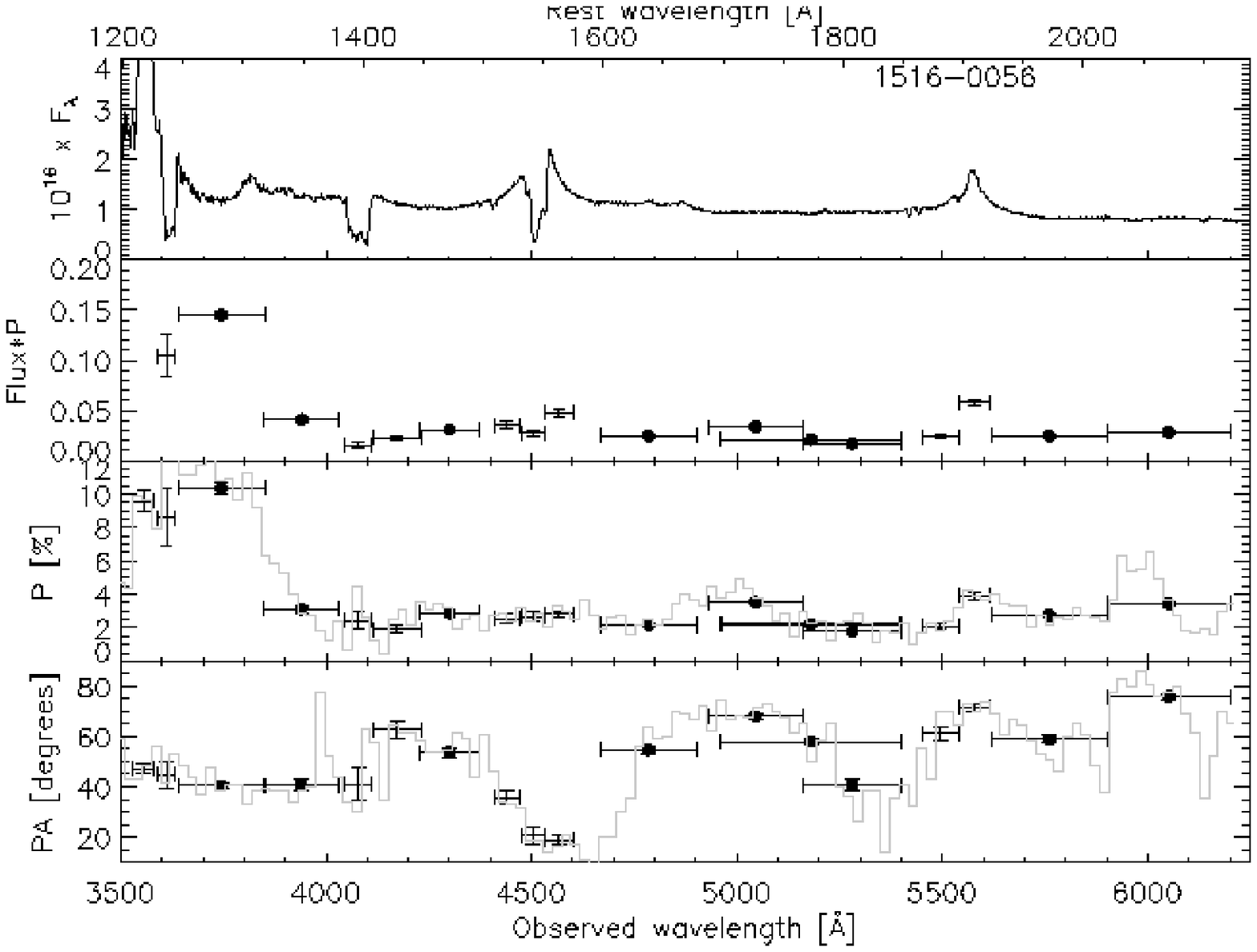}}
\end{figure}

\begin{figure}
 \ContinuedFloat
 \centering
  \subfloat[][Fig. 2$q$]{\includegraphics[width=5in]{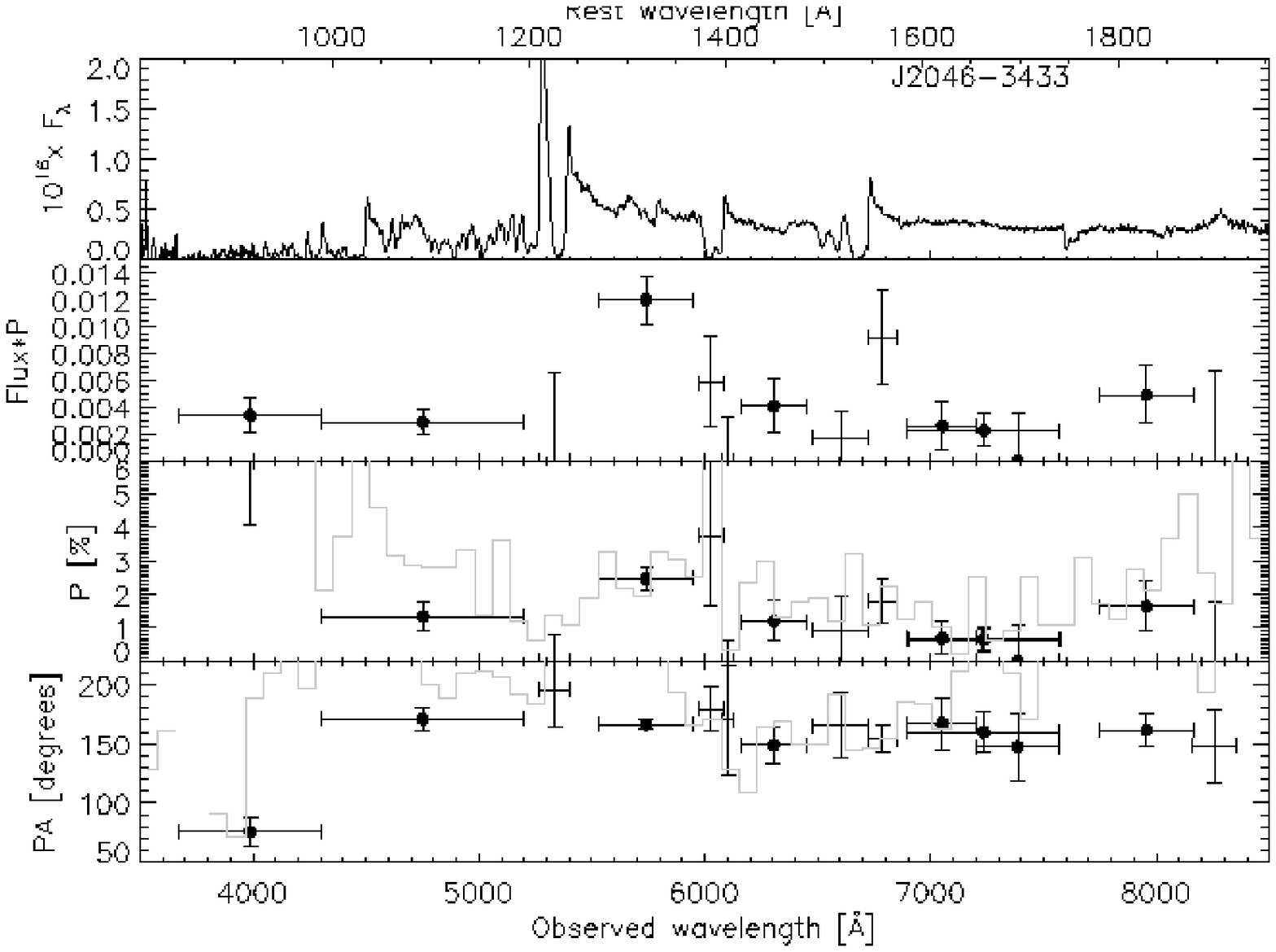}}
\end{figure}

\clearpage

\begin{figure}
 \ContinuedFloat
 \centering
  \subfloat[][Fig. 2$r$]{\includegraphics[width=5in]{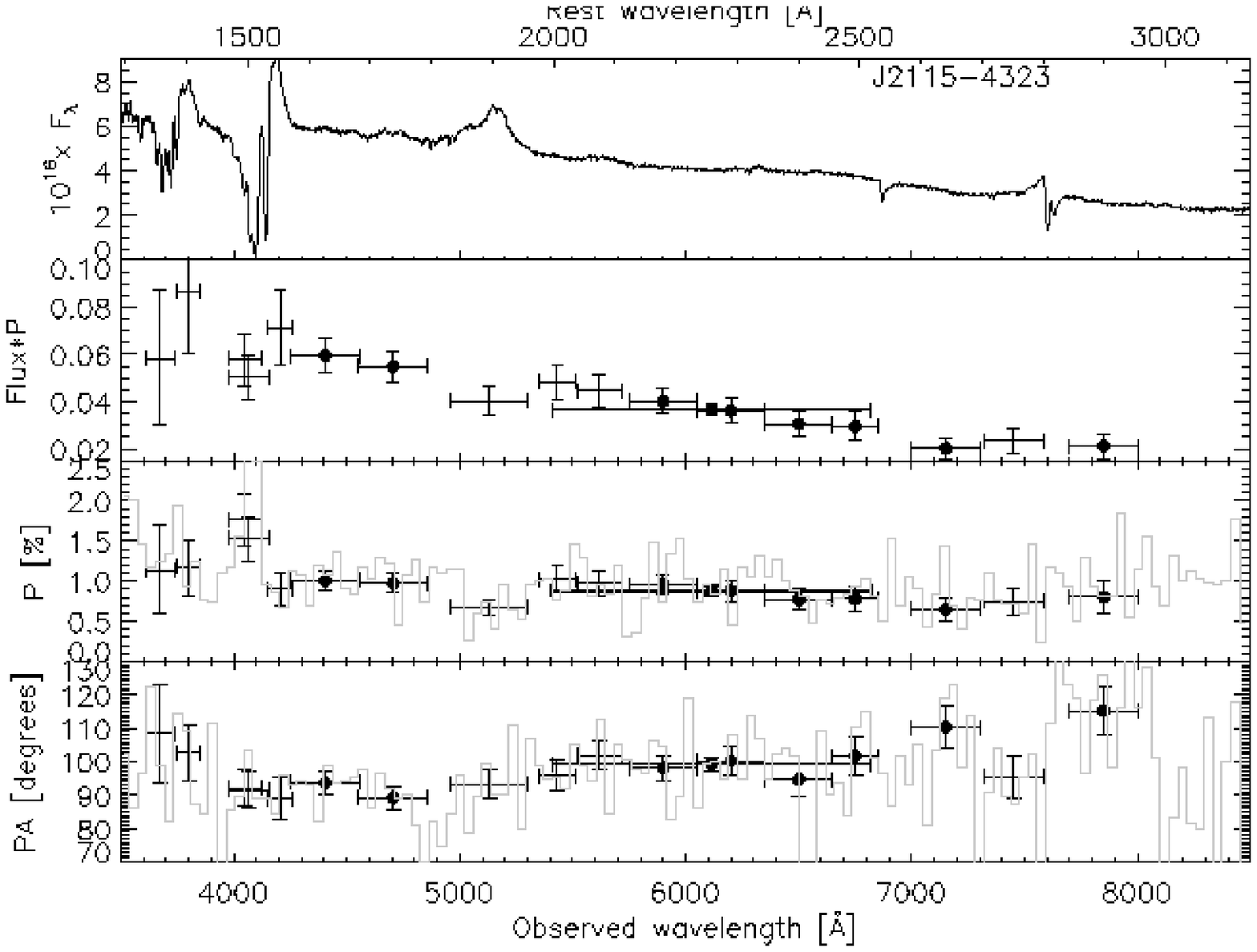}}
\end{figure}

\begin{figure}
 \ContinuedFloat
 \centering
  \subfloat[][Fig. 2$s$]{\includegraphics[width=5in]{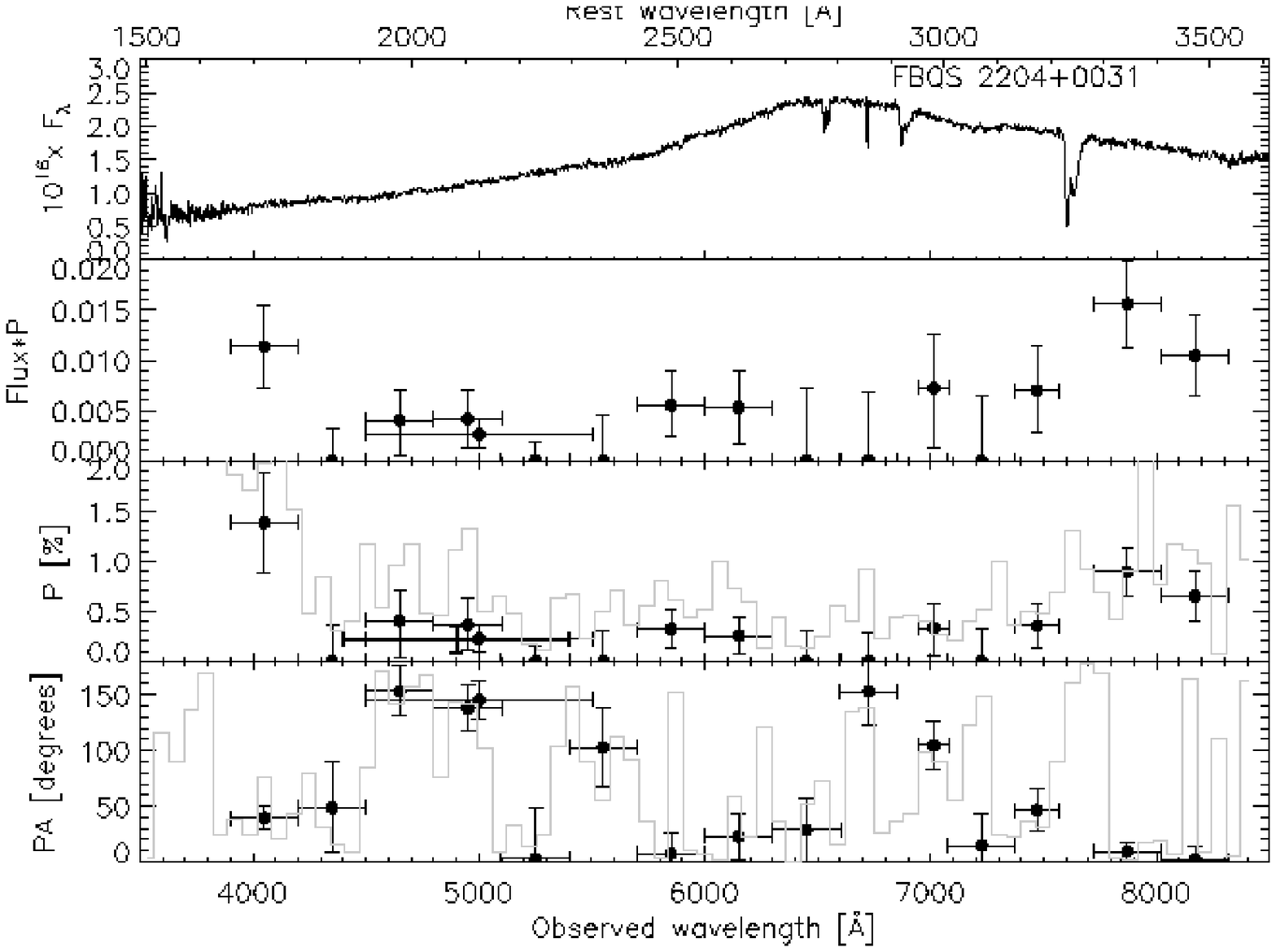}}
\end{figure}

\begin{figure}
 \ContinuedFloat
 \centering
  \subfloat[][Fig. 2$t$]{\includegraphics[width=5in]{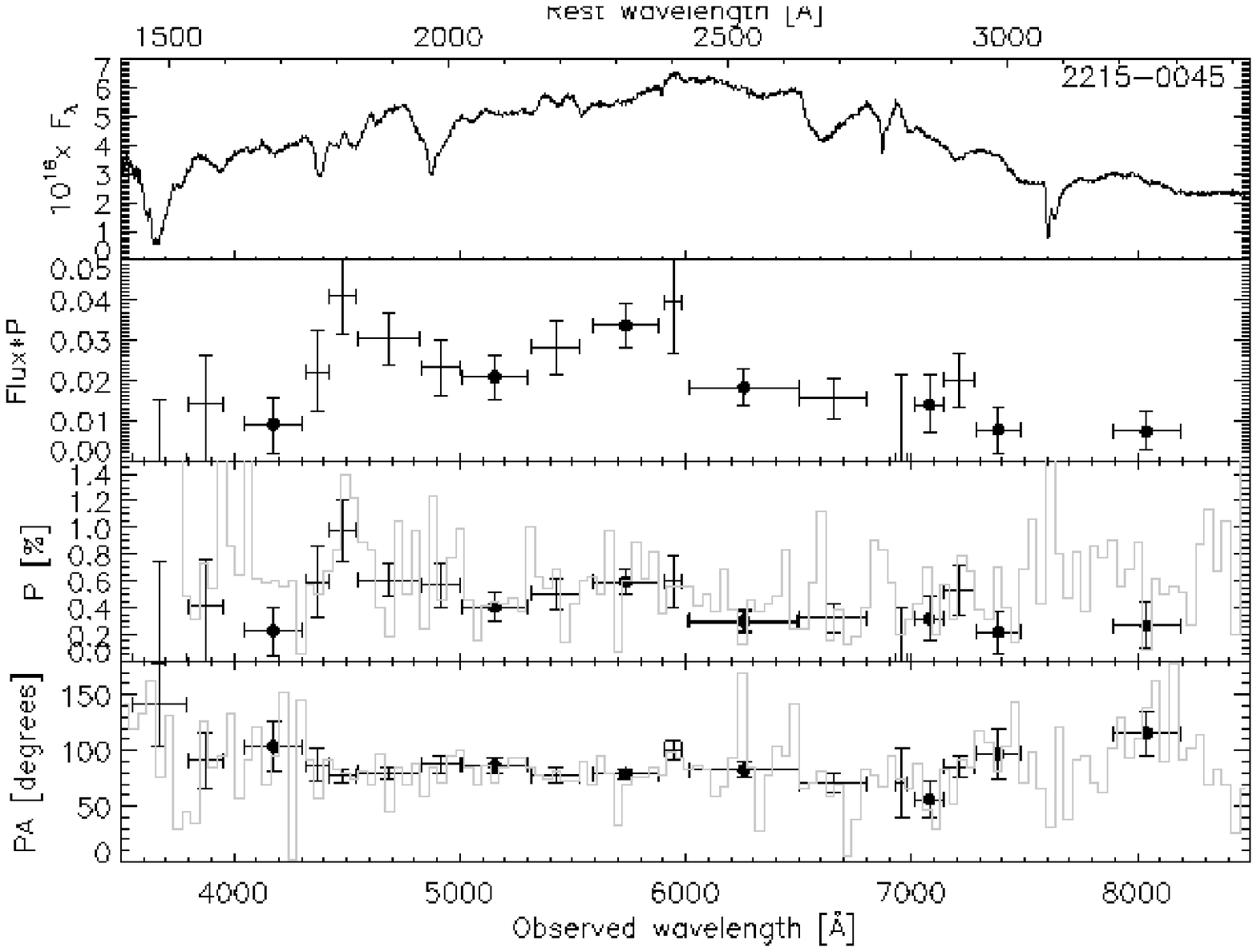}}
\end{figure}

\begin{figure}
 \ContinuedFloat
 \centering
  \subfloat[][Fig. 2$u$]{\includegraphics[width=5in]{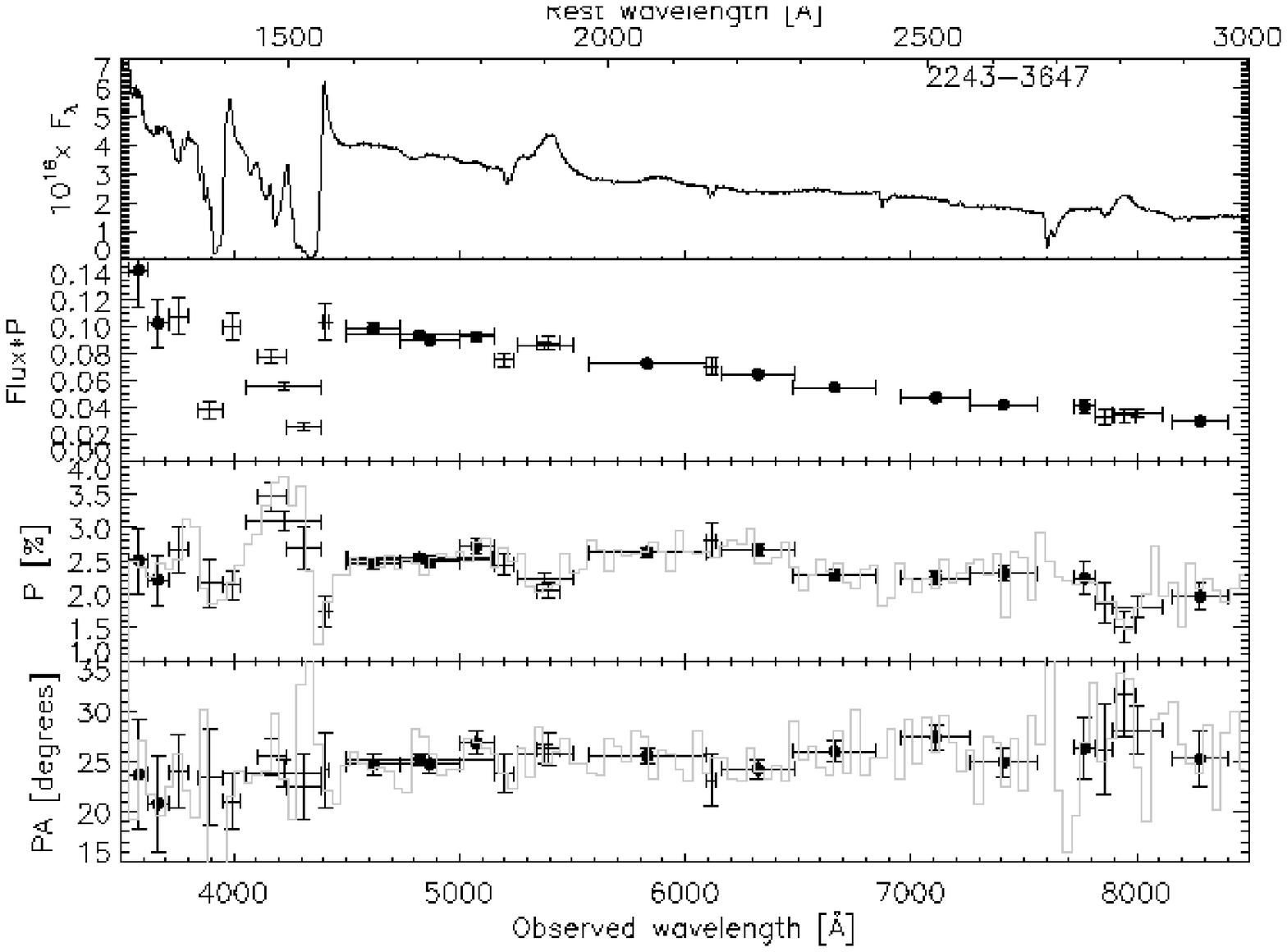}}
\end{figure}

\begin{figure}
 \ContinuedFloat
 \centering
  \subfloat[][Fig. 2$v$]{\includegraphics[width=5in]{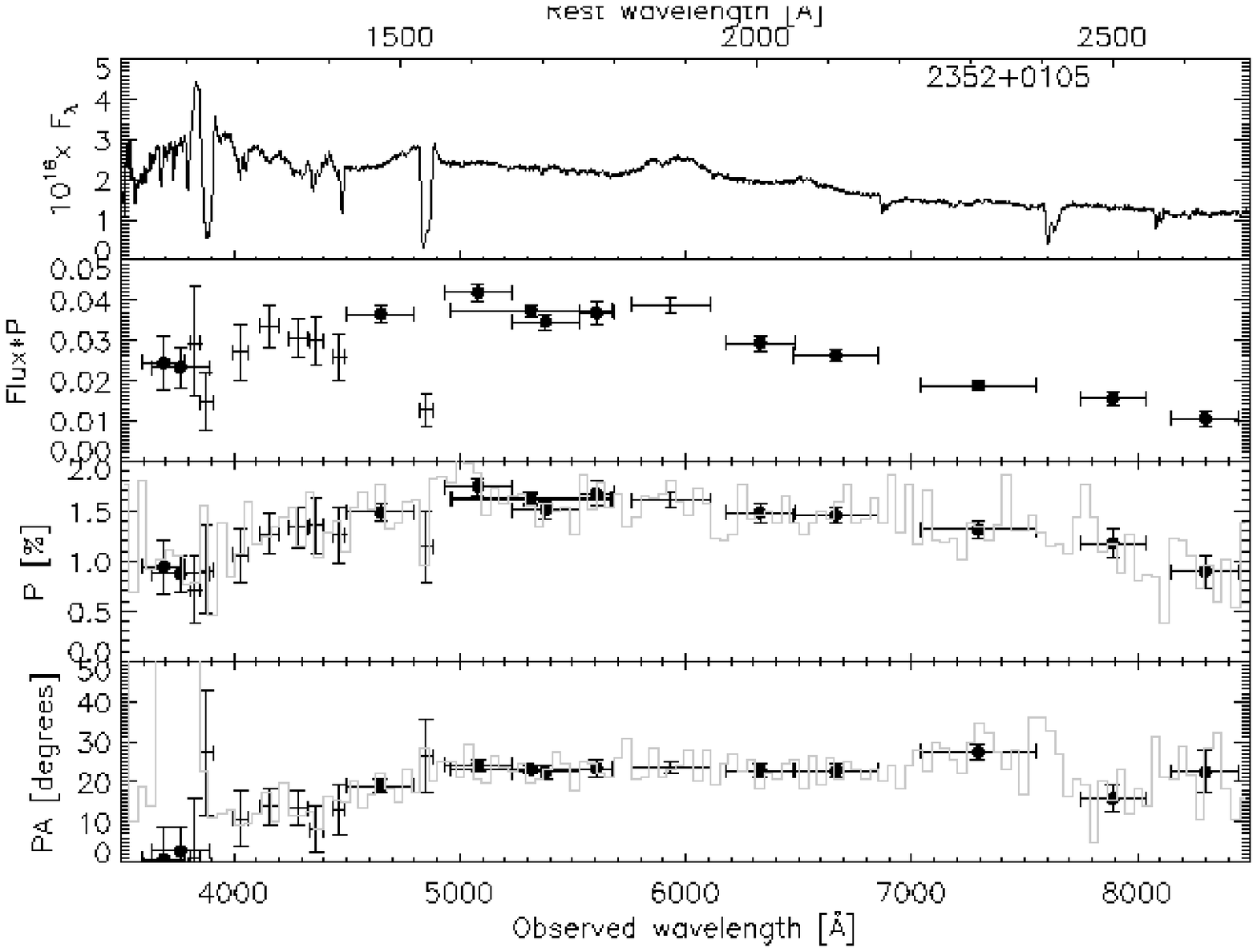}}
\end{figure}

\begin{figure}
 \ContinuedFloat
 \centering
  \subfloat[][Fig. 2$w$]{\includegraphics[width=5in]{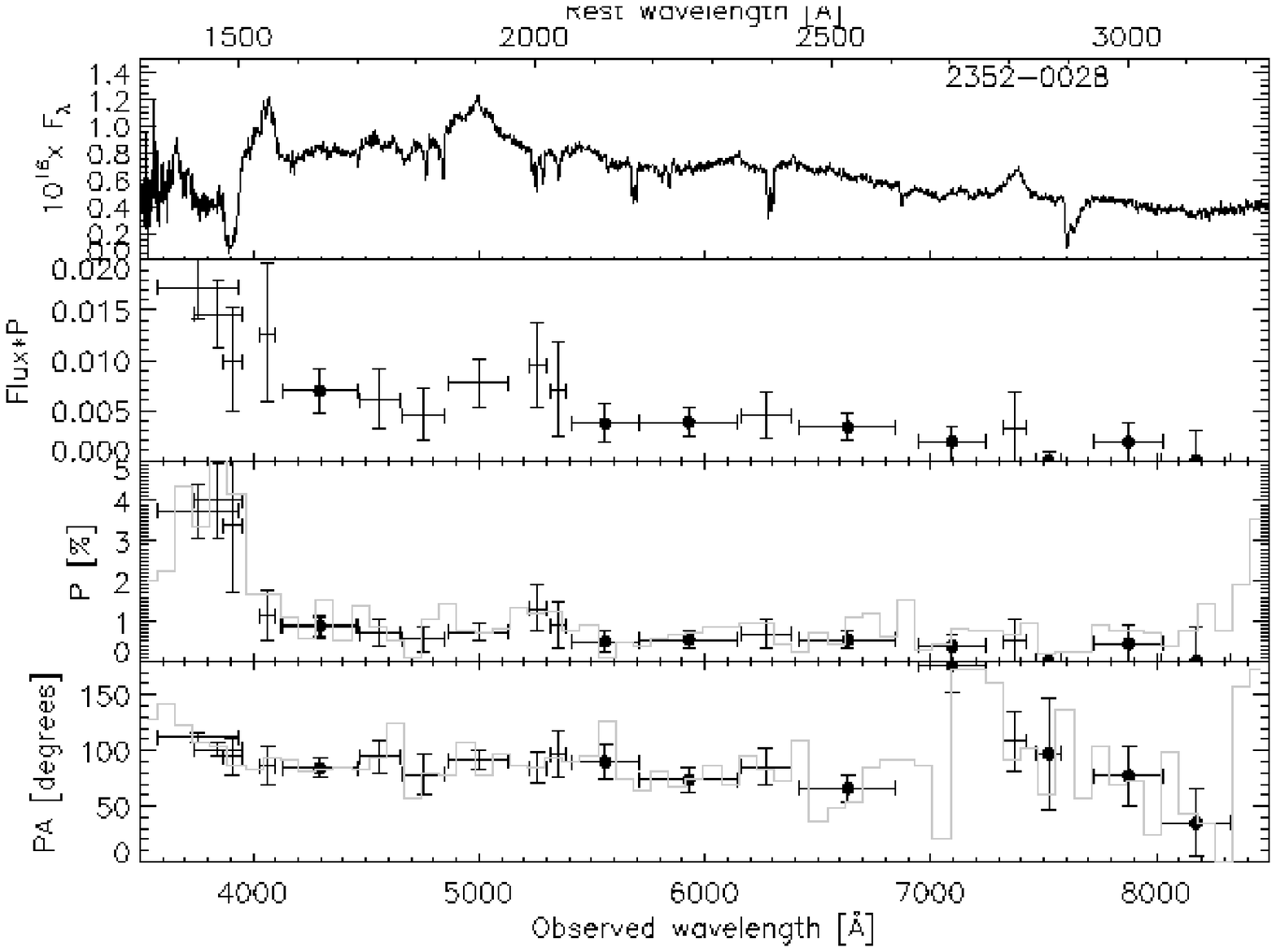}}
\end{figure}

\clearpage

\clearpage

\end{document}